\pdfoutput=1
\documentclass[letterpaper, 10 pt, journal]{IEEEtran}  
\usepackage{balance}
\usepackage{color}
\usepackage{subfigure}
\usepackage{caption}
\usepackage{cite}
\usepackage{empheq}
\usepackage{mathrsfs}
\usepackage{mleftright}

\usepackage{mathtools}

\usepackage{tikz}
\usepackage{circuitikz}
\usepackage[T1]{fontenc}

\usepackage{graphicx,amssymb,amstext,amsmath,amsthm}

\usepackage[bookmarks=false]{hyperref}

\usepackage{multirow}
\usepackage{stfloats}

\newtheorem{prop}{Proposition} 

\newtheorem{thm}{Theorem}
	
\newtheorem{lem}{Lemma}

\newtheorem{rem}{Remark}
\newtheorem{problem}{Problem}

\IEEEoverridecommandlockouts

\begin{document}

\title{\LARGE \bf Set-Point Regulation of Linear Continuous-Time Systems
using Neuromorphic Vision Sensors} 

\author{%
Prince Singh$^{\,\rm a,\star}$ \quad Sze Zheng Yong$^{\, \rm a,\star}$  \quad  Emilio Frazzoli$^{\,\rm a}$\\
\thanks{
$\star$ \ Equal contributions from authors.}
\thanks{
$^{\rm a}$ P. Singh, S.Z Yong, and E. Frazzoli are with the Laboratory for Information and Decision Systems,
Massachusetts Institute of Technology, Cambridge, MA, USA (e-mail: \{prince1,szyong,frazzoli\}@mit.edu).}
}

\maketitle
\thispagestyle{empty}
\pagestyle{empty}

\begin{abstract}
Recently developed neuromorphic vision sensors have become promising candidates for agile and autonomous robotic applications primarily due to, in particular, their high temporal resolution and low latency.
Each pixel of this sensor independently fires an asynchronous stream of ``retinal events" once a change in the light field is detected. Existing computer vision algorithms can only process periodic frames and so a new class of algorithms needs to be developed that can efficiently process these events for control tasks. 
In this paper, we investigate the problem of \textit{regulating} a continuous-time linear time invariant (LTI) system to a desired point using measurements from a neuromorphic sensor. 
We present an $H_\infty$ controller that \textit{regulates} the LTI system to a desired set-point and 
provide the set of neuromorphic sensor based cameras for the given system that fulfill the regulation task.
The effectiveness of our approach is illustrated on an unstable system. 
 \end{abstract}

\section{Introduction}

The output of a neuromorphic vision sensor is a sequence of events rather than periodic frames produced by a regular camera (e.g., CCD-, CMOS-based). We term these events as ``retinal events" since they are generated once the observed light field changes by more than a user-chosen threshold  \cite{liu.2010}.

The Dynamic Vision Sensor (DVS) is the first commercially available neuromorphic vision sensor \cite{lichtsteiner2008128} whose pixels independently and asynchronously fire retinal events once a change in the light field is detected. One big advantage of the DVS is that these retinal events are information bearing and so one avoids processing redundant data as with camera frames. 
In addition, the DVS has alluring properties, for example, micro-second temporal resolution, low-latency (order of micro-seconds) resulting in increased reactivity, high dynamic range ($>120 dB$), low power requirement, collectively making it a viable sensor for enabling the quick computation of control commands to facilitate aggressive maneuvers of agile robots. %

\emph{Literature Review.} At the current state of the art, almost all vision based control of mobile robots relies on algorithms that are developed to process the frames from `regular' cameras. These algorithms are unfortunately not suited to process the output of the low-latency neuromorphic vision sensors, which fire a sequence of asynchronous time-stamped events that describe a change in the perceived brightness at each pixel. In view of the DVS' interesting properties, this sensor seems  to be an ideal choice for tasks that are limited by sensing speed and/or sensing power; for example, tasks ranging from stabilizing the upright position of robotic insects \cite{fuller2014controlling} to enabling high speed collision-free flights of autonomous micro-aerial vehicles in complex environments   \cite{barry2015pushbroom} (not achieved yet). Other existing works use neuromorphic vision sensors for balancing an inverted pencil \cite{conradt2009pencil}, for controlling an autonomous goalie \cite{delbruck2015robotic} and for heading regulation \cite{Erich15,censi2015efficient}. 
However, all the proposed methods are problem-specific and they involve first computing explicit representations for the states and then using these estimates for closed-loop control. Hence, it remains an open problem to consider if less restrictive conditions on a given system can be achieved by going directly from the events to control commands rather than performing control via state-estimation.

Additionally, one cannot readily apply existing control techniques developed in the event-based control literature \cite{astrom2008event}, in which one typically has the flexibility to design a sensor (thus, events) to guarantee some performance requirement for the overall system (e.g., minimize the attention needed by the plant).  However, in our case, we are given a sensor and are restrained by its inherent properties (i.e., with no means of controlling the retinal events except via threshold design) to facilitate our control task. 

\emph{Contributions.}  To the best of the authors' knowledge, the prequel of this work \cite{singh.yong.ea.2016} has been the first to address the stability of a continuous time linear time invariant (LTI), single input single output (SISO) system using asynchronous 
neuromorphic measurements from a DVS. We further remark that the results of \cite{singh.yong.ea.2016} can be readily adapted for the case of multiple input single output (MISO) system, which is an artifact of the $H_\infty$ controller design procedure. The present work provides results for the case of a multiple input multiple output (MIMO) system. 
Our approach 
goes directly from the `retinal events' to control commands, instead of 
first explicitly estimating the system states for feedback control. 
The intuition behind our approach is based on characterizing the lowest upper bound on the relative error between the continuous-time output that we do not have access to and the estimate of this output computed from the retinal events fired by the DVS. 
Then, by considering an auxiliary uncertain system, we show that an $H_\infty$ controller coupled with small gain theorem
regulates the auxiliary system and in turn regulates our hybrid system to a neighborhood of a desired \emph{stabilizable} set-point within a pre-set tolerance; 
furthermore, we derive the maximum event threshold that is required for a DVS to achieve the regulation objective for the given LTI system. Our solution is facilitated with some ideas and tools drawn from works done within the context of control with limited information, in particular, the quantized control literature, e.g., \cite{elia2001stabilization,fu2005sector,vu2008stabilizing}.

\emph{Outline.} This paper is organized as follows. In Section II, we clearly formulate the problem by first characterizing the DVS model and represent the combined LTI system and DVS model as a hybrid system. 
Then, in Section III, we design a feedback controller that regulates this hybrid system to a given \emph{stabilizable} set-point. 
In turn, we present a criterion that provides us with  the least restrictive (largest) event threshold that is required of a DVS for the regulation task. 
In Section IV, we verify our results via a numerical experiment. 
Finally, in Section V, we present conclusions and outline possible extensions to 
this work.

\section{Problem Formulation} \label{sec:Problem}

\begin{figure}[t]
\centering
\includegraphics[width=0.5\linewidth]{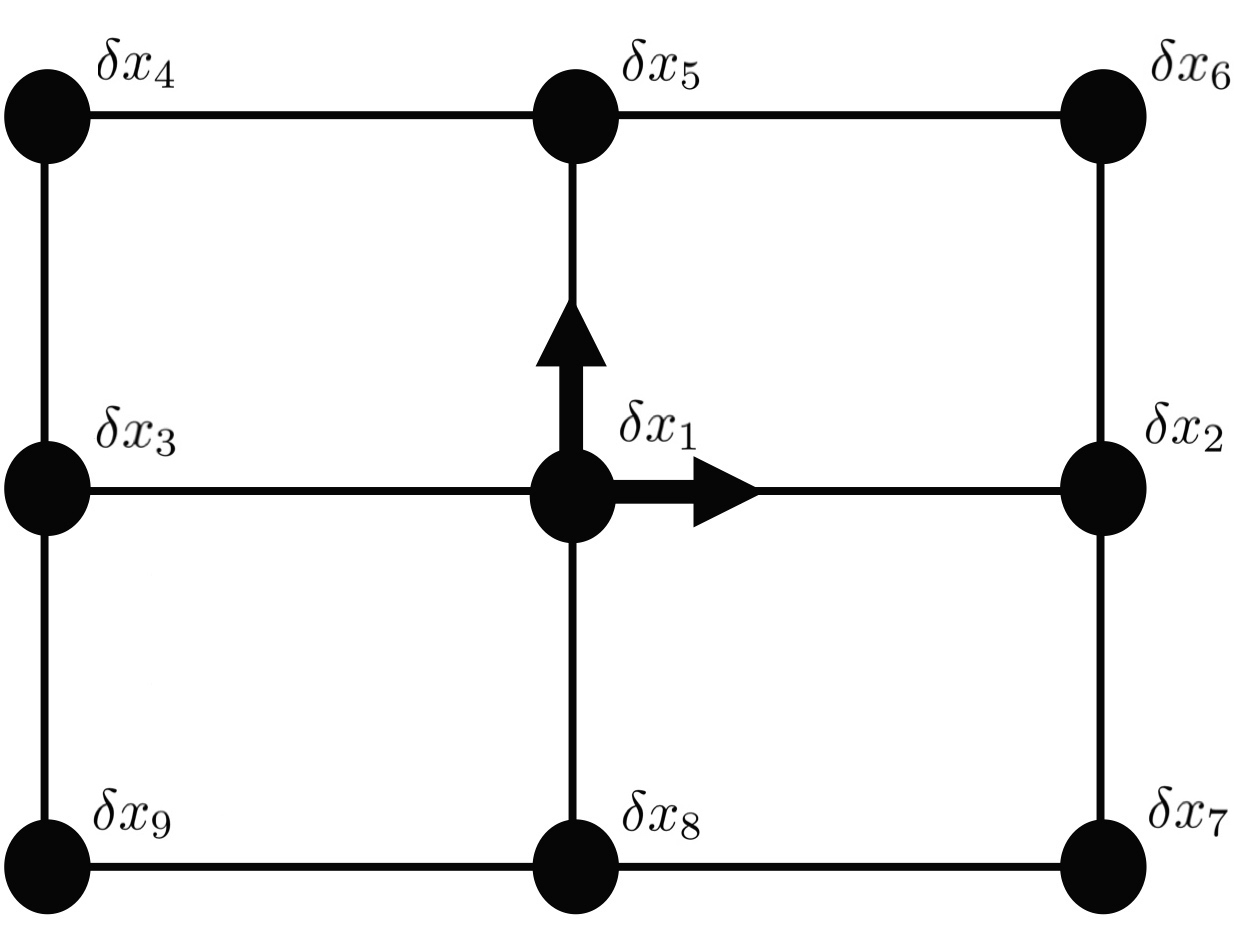} \caption{Projection of the relative locations of $r = 9$ pixels in the DVS's sensor with respect to center pixel $\delta x_1 = [0 \ 0]^T$ onto the image plane. \label{fig:grid}}
\end{figure}

\emph{LTI System.} Consider the unstable,  stabilizable  and detectable continuous time system (see Appendix \ref{sec:physicalEx} for a physical example) given by,
\begin{align} \label{eq:main1}
\begin{array}{ll}
\dot{x}^o&=A x^o+ B u^o,\\
y_i &= c'(x^o + \delta x_i)
\end{array}
\end{align}
\noindent where $x^o \in \mathbb{R}^n$ is the system's state in the original coordinate frame, $A \in \mathbb{R}^{n \times n}$, $u^o \in \mathbb{R}^{m}$ is the $m$-dimensional control input, $B \in \mathbb{R}^{n \times m}$, $c \in \mathbb{R}^{n \times 1}$, $\delta x_i \in \mathbb{R}^n$ represents the relative position of the $i$'th pixel on the DVS' sensor with respect to a reference pixel on the image plane (e.g., the center pixel $\delta x_1$) as shown in Figure \ref{fig:grid}, and $y_i \in \mathbb{R}$ for $i=1,\hdots,r$ denotes the output luminosity of the $i$'th pixel with $r$ representing the total number of the pixels that are each equipped with an independent sensor to detect brightness changes. 

From a practical application standpoint (e.g., a fleet of UAV's required to follow a fixed luminosity level to fulfill a control task), it may be desired to regulate \eqref{eq:main1} to a neighborhood of a point $x_d \in \mathbb{R}^n$ and so we consider a coordinate transformation,
\begin{align} \label{eq:coord}
\begin{array}{ll}
x^o \rightarrow x + x_d.
\end{array}
\end{align}
However, the system \eqref{eq:main1} cannot be regulated to neighborhoods of any $x_d$.  In particular, following the work of \cite{paden2015asymptotically}, the set of \textit{stabilizable} states $x_d$ are found to be those that lie in the intersection of the reachable subspace and a particular controlled invariant subspace called the constant state subspace. Hence, without loss of generality, we restrict the set-points in our problem to only the states that lie in this subspsace. Then, {in this case,} we can choose $u^o$ so that the constant $Ax_d$ resulting from the combination of \eqref{eq:main1} and \eqref{eq:coord} can be eliminated. The control input
\begin{align} \label{eq:control_coord}
\begin{array}{ll}
u^o \rightarrow u - B^\dagger A x_d,
\end{array}
\end{align}
with $B^\dagger$ being the pseudo-inverse of $B$ transforms \eqref{eq:main1} into
\begin{align} \label{eq:main}
\begin{array}{ll}
\dot{x}&=A x+ B u,\\
y_i &= c'(x+x_d + \delta x_i) 
\end{array}
\end{align}
In this work, the initial state $x(0)$ is unknown. Further, let us note that we have no direct access to the output $y_i$, except through the ``retinal event" measurements that we obtain from a neuromorphic camera, which we characterize next.

\emph{DVS Model.} Our sensor of choice is the Dynamic Vision Sensor (DVS), which is the first commercially available neuromorphic sensor \cite{lichtsteiner2008128}.  
The DVS comprises of a photodiode that converts luminosity to a photocurrent, denoted by $y_i$ as in \eqref{eq:main}, which is then amplified in a logarithmic fashion to detect brightness changes in real time. To this end, let us define the \textit{trigger condition} based on which ``retinal events" are generated by the $i$'th pixel pixel as 
 \begin{align} \label{eq:trigger}
|\tau_i| \geq h,
 \end{align}
 where 
 \begin{align} \label{eq:tau}
 \tau_i \triangleq   \log_b |y_i| -\log_b |q_i|,
 \end{align} $b$ is an arbitrary base, $q_i \in \mathbb{R}$ is the trigger reference (an internal state) of the $i$'th pixel and $h > 0$ is a user-defined event threshold for the entire camera. In the parlance of a hybrid system model, the trigger condition \eqref{eq:trigger} is a guard set, which we denote as $\mathcal{D}_i$, i.e., a ``retinal event" fires when the combined system (LTI system and DVS model for the $i$'th pixel) state $\mathbf{x_i}:=[x^\top,q_i]^\top \in \mathcal{D}_i $. 

The $k-$th ``retinal event" fired by the $i$'th pixel is given by the triple: $\langle t_k^i, \langle x_p^i(t_k^i),  y_p^i(t_k^i)  \rangle , p_i(t_k^i) \rangle$ where $t_k^i$, $  \langle x_p^i(t_k^i),   y_p^i(t_k^i) \rangle  $ and $p_i(t_k^i)$, respectively, denote the time-stamp, the pixel coordinates and polarity information whenever an event is fired. 
As aforementioned, we have no access to output $y_i$, but instead we have access to polarity measurements, $p_i \in \{-1,0,+1\}$ given by the events:
\begin{align} \label{content}
\begin{array}{ll}
p_i &=\left\{\begin{array}{ll}
\textrm{sgn}( \tau_i), & {\textrm{if }}  \mathbf{x_i} \in \mathcal{D}_i,\\
0, & \rm{otherwise}.
\end{array}\right.
\end{array}
\end{align}

\noindent Due to the continuity of the output trajectory $y_i$, we know that the event triggering for the DVS always takes place when equality holds for the trigger condition \eqref{eq:trigger}. Thus, the evolution of the trigger reference is described by
\begin{align} \label{content1}
\begin{array}{ll}
{q}^+_i&=q_i \rho^{-p_i},
\end{array}
\end{align}

\noindent where for convenience, we define 
\begin{align} \label{eq:rho}
\rho \triangleq b^{-h} \in (0,1),
\end{align} 
 as the spacing of the logarithmic partitions induced onto the output space  
 by the logarithmic \textit{trigger condition} in \eqref{eq:trigger}. 
This choice of $\rho$ in \eqref{eq:rho} captures the range of positive values for the event threshold $h$ of the DVS. Then, the \textit{trigger condition} \eqref{eq:trigger} can be equivalently re-written as
\begin{align} \label{eq:guardSet}
\mathcal{D}_i\triangleq \{\mathbf{x_i} : \left|\frac{y_i}{q_i}\right| \geq \frac{1}{\rho}~ \textrm{or} ~ \left|\frac{y_i}{q_i}\right| \leq \rho\},
\end{align}
which explicitly defines our guard set $\mathcal{D}_i$ in terms of $\rho$.

 Further, we make the following assumption regarding the trigger reference $q_i$:
 \begin{description} 
 \item[(A1)] The initial trigger reference $q_i(0)$ lies in the interval: $m_i \leq q_i(0) \leq M_i$ where $0 < m_i \leq M_i$ are scalars and satisfies $\rho< \left| \frac{y_i(0)}{q_i(0)} \right|<\frac{1}{\rho}$ 
 with $y_i(0)\in \mathcal{D}_i^c$, the complementary of the guard set $\mathcal{D}_i$.
 \end{description}
This assumption is more realistic than the assumption made in our previous work \cite{singh.yong.ea.2016} in that the luminosity of the environment that a DVS is turned on in may not be known exactly, but is only known to be within some bound. 
Intuitively, this assumption along with the continuity of the output trajectory 
makes it possible to keep track of the internal trigger reference at all times. 
Since the luminosity is always positive, as enforced in (A1), the sign of $q_i$ is known at all times, which also relaxes the assumption made in \cite{singh.yong.ea.2016}. 



\begin{figure}[t]
\centering
\includegraphics[width=0.7\linewidth]{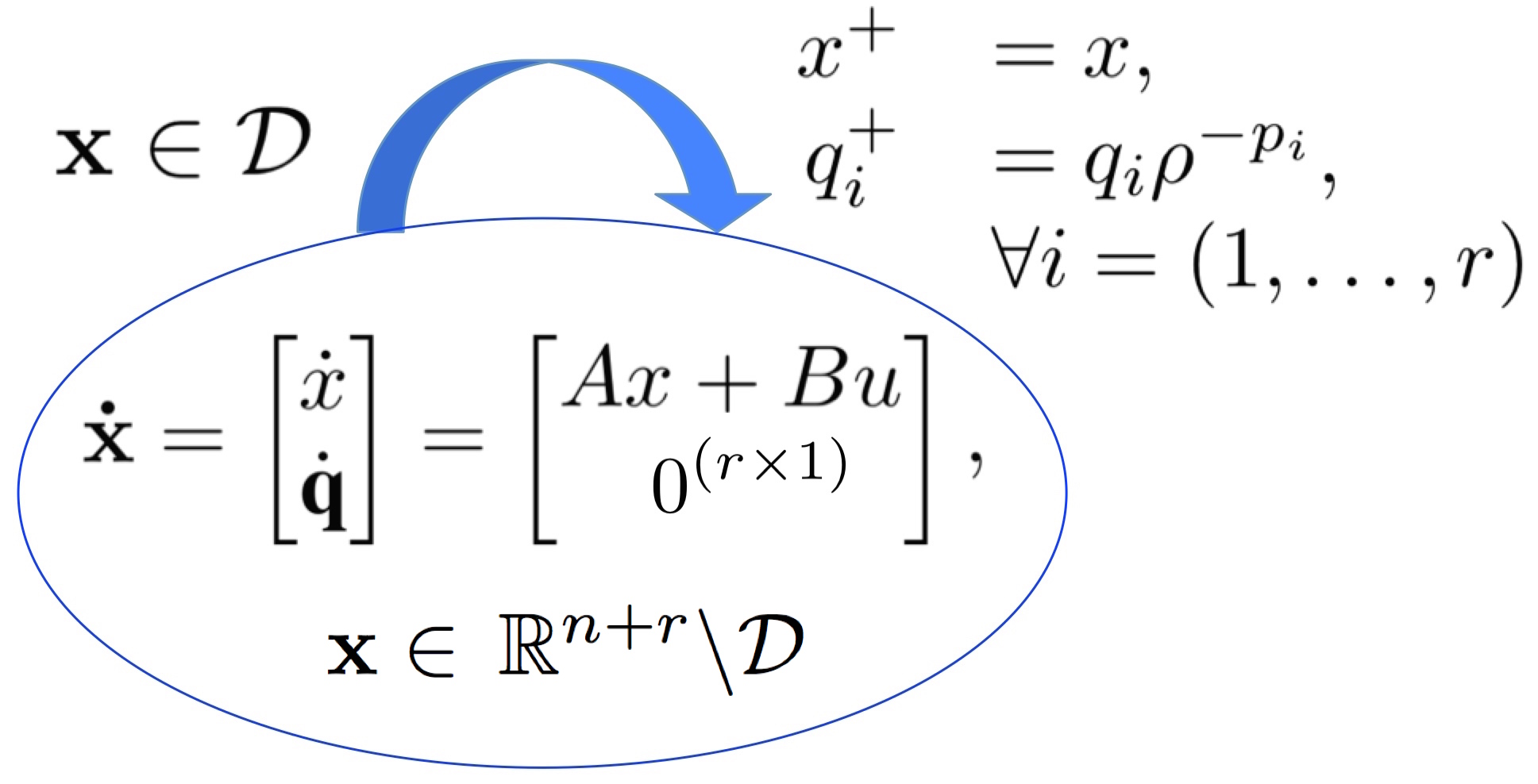} \caption{Open loop hybrid automaton of combined LTI system and DVS model in \eqref{comb}, where $\mathcal{D} = \cup_{i=1}^r\mathcal{D}_i$ and $\mathcal{D}_i$ is defined in \eqref{eq:guardSet}. \label{fig:hybridAutomaton}}
\end{figure}

\emph{Combined System.} Now, combining the LTI system in \eqref{eq:main} and DVS model in \eqref{content} with $r$ sensors yields the following hybrid system:
\begin{align} \label{comb}
\begin{array}{rlll}
\mathbf{\dot{x}}&=\begin{bmatrix}\dot{x}\\ \dot{\textbf{q}} \end{bmatrix}&=\begin{bmatrix} Ax+Bu \\ 0^{(r \times 1)} \end{bmatrix}, & \mathbf{x} \in \, \mathbb R^{n+r} \textbackslash
\mathcal D, \,\\[0.35cm] 
\mathbf{\dot{x}}^+ &= \begin{bmatrix}x^+\\{\textbf{q}}^+\end{bmatrix}&=
\begin{bmatrix} x\\ \begin{bmatrix}
q_1 \rho^{-p_1} \\ 
\vdots \\
q_r \rho^{-p_r}
\end{bmatrix} \end{bmatrix}, & \mathbf{x} \in \mathcal{D}, \,
\\
\end{array}
\end{align}
where ${\textbf{q}} \triangleq [q_1,\dots,q_r]^T$, the polarity measurement $p_i$ for the $i$'th pixel is given in \eqref{content} and with a slight abuse of notation, we denote the set of disjoint pixel-wise guard sets $\mathcal{D}_i$ defined in \eqref{eq:guardSet} with the guard set of this hybrid system $\mathcal{D}$. The hybrid automaton that results is illustrated in Figure \ref{fig:hybridAutomaton}. \\

Now, the regulation control problem with the DVS reads:
\begin{problem}\label{problem}
\noindent The \emph{objective} of this paper is two-fold: 
\begin{enumerate}
\item Design an appropriate feedback controller $u$   
that incorporates polarity measurements given by a DVS to regulate the hybrid system \eqref{comb} to an $\epsilon$-neighborhood of a \textit{stabilizable} state $x_d$, i.e., guaranteeing for a given tolerance/precision $\epsilon > 0$, $\lim_{t \rightarrow \infty} ||x^o(t) - x_d||_2 = \lim_{t \rightarrow \infty} ||x(t)||_2 < \epsilon$. 
\item For the controller designed for Problem \ref{problem}-1, find the least restrictive (largest) upper-bound on the event threshold $h^*$, such that for any DVS with event threshold $h<h^*$, this controller regulates the hybrid system \eqref{comb} to an $\epsilon$-neighborhood of the \textit{stabilizable} state $x_d$.
\end{enumerate}
\end{problem}

\section{Controller Design} \label{sec:controller}

\begin{figure}
  \subfigure[Feedback controller in closed loop with 
  $y_i, \forall i = \{1, \dots,r \}$ is defined in \eqref{eq:main} and
  $\mathbf{p} \triangleq {[}p_1,\dots,p_r{]}^T$, whose entries are given in \eqref{content}.]{
  \includegraphics[width=\linewidth]{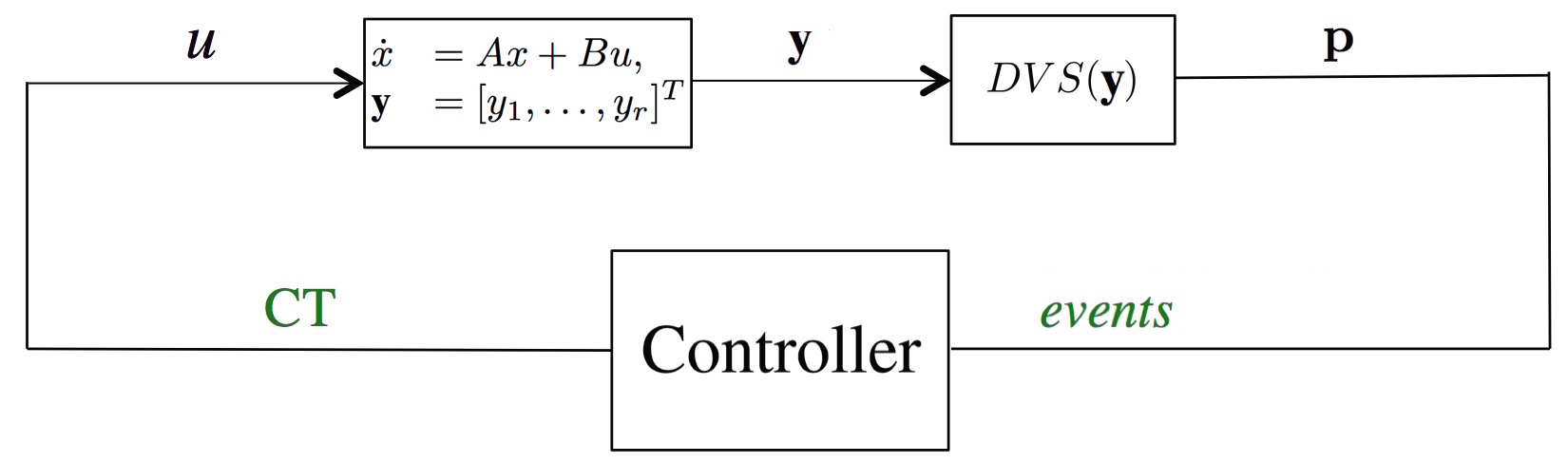}  \label{approach}}

  \subfigure[Cascade decomposition of the feedback controller. ]{
  \includegraphics[width=\linewidth]{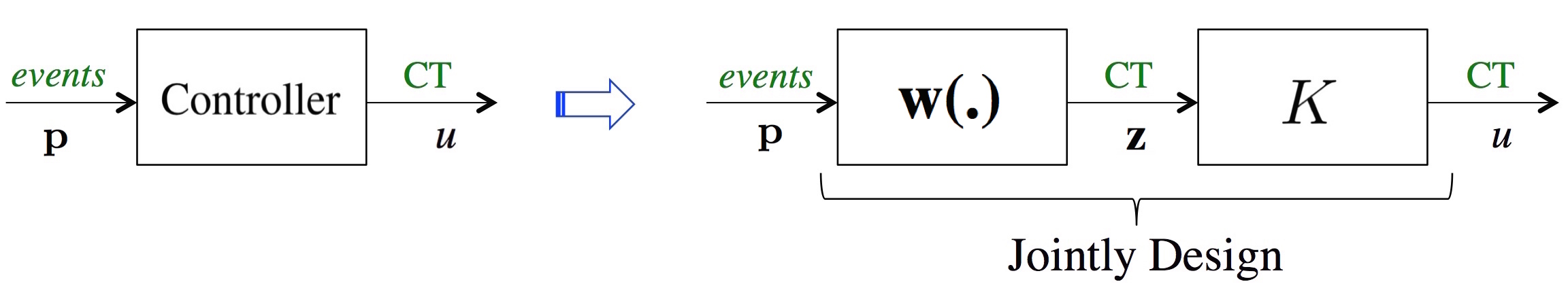}   \label{feedback}}
  
    \subfigure[Decomposition of $w_i(.)$ function. ]{
    \includegraphics[width=\linewidth]{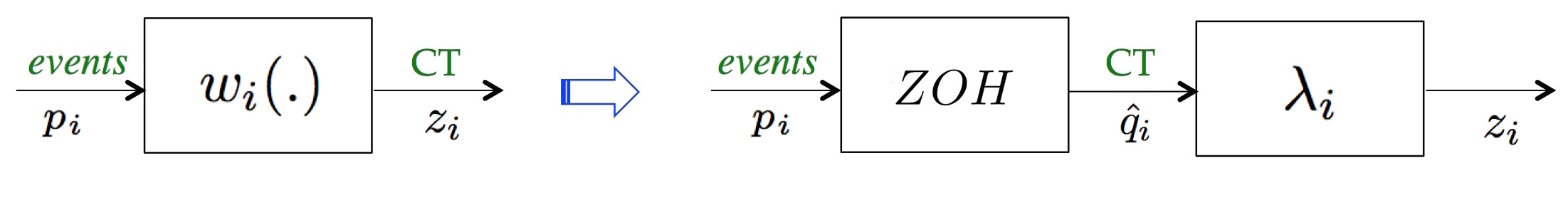}   \label{wfunc}}

  \caption{Controller design approach: From asynchronous events to continuous-time (CT) control command. }
\end{figure}

In this work, we propose a controller that uses polarity measurements from all pixels of the DVS in \eqref{content} to regulate the LTI system \eqref{eq:main}. 
As shown in Figure \ref{approach}, the LTI system outputs a continuous time signal $\textbf{y} = [y_1,\dots,y_r]^T$, whose 
respective pixel of the DVS produces retinal events based on the \textit{trigger condition} in \eqref{eq:trigger}. Our goal is to design a feedback controller that operates on the incoming events to generate a continuous time control signal $u$ that would regulate the pair $(A,B)$ within a neighborhood of a desired set-point. 

The intuition behind our controller design is based on isolating the uncertainty in the pixel-wise polarity measurements of the DVS. Inspired by output feedback control, we construct an estimator $\textbf{w(.)} = [w_1(.),\dots,w_r(.)]^T$ for the output signal $\textbf{y}$ in Section \ref{sec:estimator} and use the resulting estimate $\textbf{z} = [z_1(t),\dots,z_r(t)]^T$ as an input to our controller $K$, which we will design in Section \ref{sec:controller}. 
This cascade set-up of the feedback controller, which consists of the estimator $\textbf{w(.)}$ and the controller $K$ is shown in Figure \ref{feedback}.
%
%




\subsection{Design of estimator for $\textbf{y}$} \label{sec:estimator}

Since we have no measurements between retinal events, a relatively simple design for each component of $\textbf{w(.)}$, $w_i(.)$, may be one that performs a Zero-Order-Hold (ZOH) on the retinal events arriving from the DVS that is then amplified by a non-zero scalar $\lambda_i$ as shown in Figure \ref{wfunc}.  
Thus, we will construct the signal $z_i$ as an estimate of $y_i$ with the following:
\begin{align}\label{y-observer}
\begin{array}{ll}
z_i&=\lambda_i  \hat{q}_i, \ \ 0 \neq \lambda_i \in \mathbb{R},
\end{array}
\end{align}
where $\lambda_i$ will be provided in Section \ref{sec:analysis} and $\hat{q}_i$ is an estimate of $q_i$. 
It is clear that the entries of the vector \textbf{z} comprises of \eqref{y-observer}. 
We consider estimates of the trigger reference $q_i$ that follow the same evolution of \eqref{content1}: 
\begin{align}\label{q-observer}
\hat{q}_i^+ = \hat{q}_i \rho^{-p_i},
\end{align}
where $\hat{q}_i(0)$ will be characterized in the following lemma. 

\begin{lem}
\label{qhat-estimator-lem2}
The following choice for the relationship between the initial conditions of the estimate of the trigger reference $\hat{q}_i(0)$ and the unknown trigger reference $q_i(0)$ 
\begin{equation}\label{qhat-estimator-init}
\hat{q}_i(0) = (1+\Delta_q^i)q_i(0), \ |\Delta_q^i| \leq \delta_q^i.
\end{equation}
\noindent is the least restrictive if
$$\hat{q}_i(0) = \frac{2m_iM_i}{M_i+m_i},$$ 
\noindent which produces the minimal $|\Delta_q^i|$ with bounds,
$$\delta_q^i = \frac{M_i-m_i}{M_i+m_i},$$
where $0 \leq \delta_q^i < 1$.
\end{lem}

\begin{IEEEproof} 
In view of $m_i \leq {q}_i(0) \leq M_i$, according to (A1),
$$-M_i \leq -{q}_i(0) \leq -m_i ,$$
then, 
$$ \frac{\hat{q}_i(0)-M_i}{{q}_i(0)} \leq \frac{\hat{q}_i(0)-{q}_i(0)}{{q}_i(0)} \leq \frac{\hat{q}_i(0)-m_i}{{q}_i(0)},$$
for some $q_i(0) > 0$ (i.e., luminosity is positive) and further bounding the above with respect to the boundary values of ${q}_i(0)$, we have,
\begin{equation}\label{q_vals_bound}
\frac{\hat{q}_i(0)-M_i}{M_i} \leq \frac{\hat{q}_i(0)\hspace{-0.05cm}-\hspace{-0.05cm}M_i}{{q}_i(0)} \leq \Delta_q^i  \leq \frac{\hat{q}_i(0)\hspace{-0.05cm}-\hspace{-0.05cm}m_i}{{q}_i(0)} \leq \frac{\hat{q}_i(0)-m_i}{m_i},
\end{equation}
where upon re-arranging \eqref{qhat-estimator-init}, we have 

$$\Delta_q^i = \frac{\hat{q}_i(0)-{q}_i(0)}{{q}_i(0)}.$$

\noindent Then, the least restrictive relationship between $\hat{q}_i(0)$ and $q_i(0)$ in \eqref{qhat-estimator-init}
is found as the minimum value of $|\Delta_q^i|$. More precisely, by equating the upper bound of \eqref{q_vals_bound} with the negative of its lower bound, we obtain 
$$\hat{q}_i(0) = \frac{2m_iM_i}{M_i+m_i},$$ which in turn, results in the minimal $|\Delta_q^i|$ as

$$\delta_q^i =  \frac{\hat{q}_i(0)-M_i}{M_i} = \frac{\hat{q}_i(0)-m_i}{m_i} = \frac{M_i-m_i}{M_i+m_i},$$
with $0 \leq \delta_q^i < 1$ since we assumed that $0 < m_i \leq M_i$.
\end{IEEEproof}

The following lemma presents the estimator for $\hat{q}_i(t)$. 
\begin{lem}
\label{qhat-estimator-lem}
The following estimator characterizes the evolution of the estimate of the trigger reference $\hat{q}_i(t)$ in a least restrictive fashion,
\begin{equation}\label{qhat-estimator}
\hat{q}_i(t) = (1+\Delta_q^i)q_i(t), \ |\Delta_q^i| \leq \delta_q^i,
\end{equation}
\noindent where $\hat{q}_i(0)$ and $\delta_q^i$ are given in Lemma 1.
\end{lem}

\begin{IEEEproof} 
Under the choice \eqref{qhat-estimator-init} in Lemma 1,
$$\hat{q}_i(0) = (1+\Delta_q^i)q_i(0),$$

\noindent and noting that the dynamics of ${q}_i^+,\hat{q}_i^+$ given by \eqref{content1} and \eqref{q-observer}, respectively, at transition are identical, then, it holds for all $t$ that
$$\hat{q}_i(t) = (1+\Delta_q^i)q_i(t),$$
since $\Delta_q^i$ takes values in a finite set and is a constant at any time.
\end{IEEEproof}

\begin{rem}\label{estimator_choice}
The choice of $\hat{q}_i(0),\delta_q^i$ in \eqref{qhat-estimator} yielding the least restrictive $z_i$ in \eqref{y-observer}
will be verified in Theorem \ref{thm:lambda} in Section \ref{sec:controller}.
\end{rem}





\begin{figure}
\centering
  \includegraphics[width=3.5in]{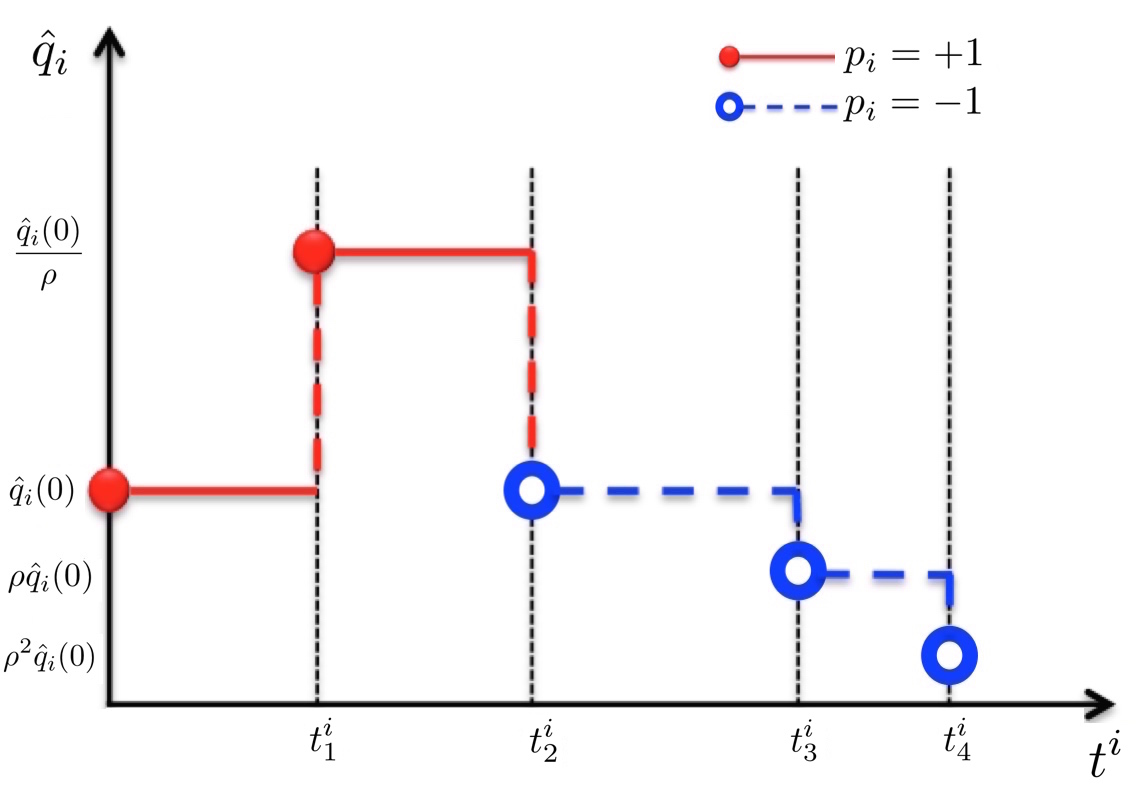}
  \caption{Output of ZOH function (estimator of $\hat{q}_i$): red/solid dots indicate positive transitions while blue/hollow dots indicate negative transitions. }
  \label{trans}
\end{figure}

Figure \ref{trans} illustrates an example scenario for the evolution of \eqref{q-observer}, 
where the brightness increased (i.e., given by the red/solid dots) at the event-time $t_1^i$ for all $q_i(0)$ in (A1) satisfying $\rho< \left| \frac{y_i(0)}{q_i(0)} \right|<\frac{1}{\rho}$ 
and the brightness decreased (i.e., given by the blue/hollow dots) at event-times $t_2^i, t_3^i, t_4^i$. It is clear that the ZOH characterizes the information between the retinal events as there is no additional additional sensor to appropriately quantify this lack of information. 


%
%

\subsection{Error quantification} \label{sec:analysis}

\begin{figure}
\centering
  \includegraphics[width=1\linewidth]{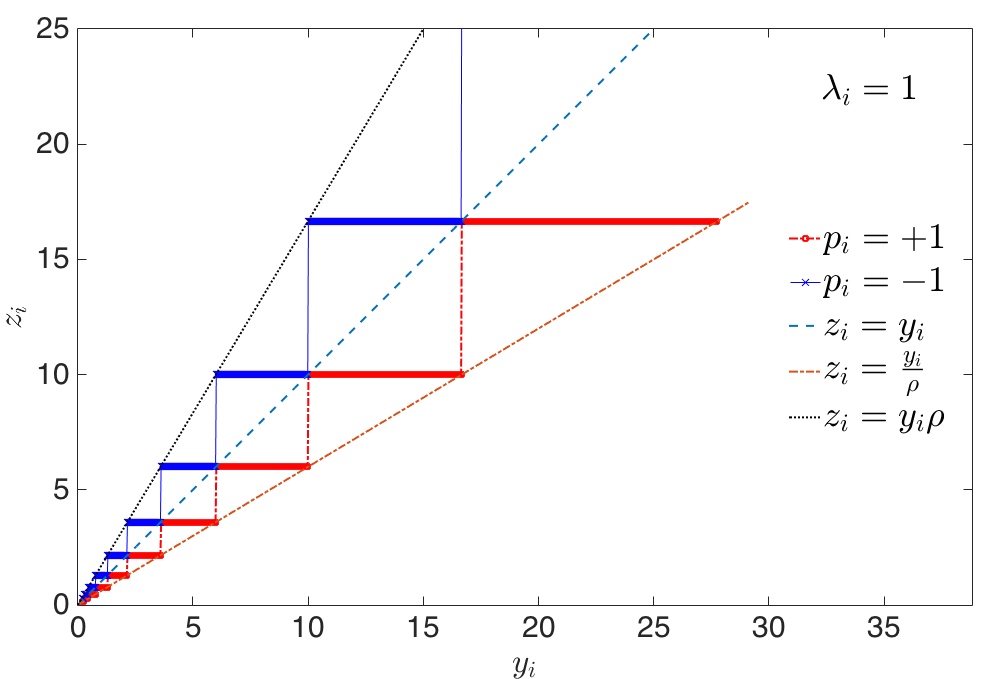}
  \caption{Analysis of the signal $z_i$ ($\lambda_i=1$): Red/solid lines are due to positive transitions ($p_i=+1$) while blue/solid-star lines are due to negative transitions ($p_i=-1$). \label{sector}}
\end{figure}

Now, we would like to quantify the closeness of the designed continuous-time signal $z_i$ to the unknown output $y_i$ of the plant. More precisely, we would like to ascertain this closeness in the sense that the maximum absolute relative error $\left|\frac{z_i-y_i}{y_i}\right|$ is minimized for each of the $r$ pixels.
%
The closeness between $z_i$ and $y_i$ can be visualized using Figure \ref{sector}; 
thus, we have 
 the following lemma.

%

\begin{lem}
\label{inequality-lem}
$\hat{q}_i$ estimates $y_i$ with bounded (asymmetric) uncertainty:
\begin{equation}\label{inequality}
\rho y_i \le q_i  = \frac{\hat{q}_i}{1+\Delta_q^i}  \le \frac{y_i}{\rho}.
\end{equation}
\end{lem}

\begin{IEEEproof} 
By the definition of the complementary guard set $\mathcal{D}_i^c$ (i.e., when a ``retinal event" doesn't fire), with the guard set $\mathcal{D}_i$ defined in \eqref{eq:guardSet}, we have that 
$$\rho y_i  < q_i < \frac{y_i }{\rho},$$ 
is assumed to hold initially.  

At an arbitrary time $t$, assume that the above inequality holds. Then, until a ``retinal event" fires, the above inequality still holds. 
Immediately before a transition occurs due to a ``retinal event" firing, we have $y_i  = \rho^{-p_i } q_i $ by the definition of the trigger condition in \eqref{eq:trigger}. After reset, we have $y_i ^+ = y_i $ and $q_i ^+ = \rho^{-p_i }q_i $ by the continuity of $y_i$, thus, it follows that $y_i ^+ = q_i ^+$ and so $$\rho y_i  \le q_i  \le \frac{y_i }{\rho}$$ holds after the transition. By induction, $\rho y_i \le q_i \le \frac{y_i}{\rho}$ holds at all times. Finally, by Lemma \ref{qhat-estimator-lem}, we have 
$$q_i (t) = \frac{\hat{q}_i (t)}{ 1+\Delta_q^i} $$ 
holds for all time $t$, then, the previous inequality becomes 
$$\rho y_i \le q_i  = \frac{\hat{q}_i}{1+\Delta_q^i}  \le \frac{y_i}{\rho}, $$
which yields our desired result. 
\end{IEEEproof}
Note that Figure \ref{sector} has been generated with unit amplification, $\lambda_i=1$, i.e., $z_i=\hat{q}_i$.  We observe that the estimate $\hat{q}_i$ causes an unequal spacing between positive ($p_i=+1$) and negative ($p_i=-1$) transitions (note the unequal length of the blue and red segments). 
Furthermore, the logarithmic property of the \textit{trigger condition} in \eqref{eq:trigger} enables us to conservatively bound the error between the signals $z_i$ and $y_i$ via a sector whose borders are represented by the lines $z_i = \frac{y_i}{\rho}$ and $z_i = y_i \rho$. The ability to bound this error will facilitate the ensuing analysis in designing $K$ in Section \ref{sec:controller}. Moreover, it is noteworthy that the uncertainty in our problem is similar but not equivalent to the uncertainties encountered in logarithmically quantized systems because of the `overlap between partitions' that results from the possibility for positive and negative transitions.

The following lemma provides a symmetric bound on the absolute relative error between the $z_i$ and $y_i$ signals; this symmetry is a desired trait as will be shown in Theorem \ref{thm:lambda}. 


\begin{lem}
\label{SectorError}
The upper-bound $\delta_z^i$ on the absolute relative error between the $z_i$ and $y_i$ signals 
is given by,
$$\left| \frac{z_i-y_i}{y_i} \right|\leq  \delta_z^i,$$
with 
$$z_i=\lambda_i \hat{q}_i, $$
where $\delta_z^i \triangleq \frac{M_i -m_i \rho^2}{M_i +m_i \rho^2}$ and $\lambda_i  \triangleq \frac{(M_i +m_i) \rho}{M_i +m_i \rho^2}$.
\end{lem}


\begin{IEEEproof} 
From \eqref{inequality} in Lemma 3,
$$ \rho y_i \le \frac{\hat{q}_i}{1+\Delta_q^i} \le \frac{y_i}{\rho},$$
\begin{equation}\label{bounds_del}
\min_{\Delta_q^i} (1+\Delta_q^i) \rho y_i  \le \hat{q}_i \le \max_{\Delta_q^i} (1+\Delta_q^i) \frac{y_i}{\rho},
\end{equation}
and bounding the above in view of $\delta_q^i$, given in Lemma \ref{qhat-estimator-lem2}, which produces the minimal $|\Delta_q^i|$, we have
\begin{align*}
\begin{array}{ll}
(1-\delta_q^i) \rho y_i &\le \min_{\Delta_q^i} (1+\Delta_q^i) \rho y_i  \le \dots \\
& \dots \hat{q}_i  \le \max_{\Delta_q^i} (1+\Delta_q^i) \frac{y_i}{\rho} \le (1+\delta_q^i) \frac{y_i}{\rho},
\end{array}
\end{align*}
 $$ \frac{2m_i \rho^2}{M_i\hspace{-0.05cm}+\hspace{-0.05cm}m_i\rho^2} y_i  \le z_i = \lambda_i \hat{q}_i =  \frac{(M_i+m_i) \rho}{M_i+m_i\rho^2} \hat{q}_i \le \frac{2M_i}{M_i \hspace{-0.05cm}+\hspace{-0.05cm}m_i\rho^2} y_i,$$
$$ \left(1 - \frac{M_i-m_i\rho^2}{M_i+m_i\rho^2} \right) y_i \le z_i \le \left(1+\frac{M_i-m_i\rho^2}{M_i+m_i\rho^2}\right) y_i, $$
$$ (1-\delta_z^i)y_i \le z_i \le (1+\delta_z^i)y_i,$$
which is a symmetric inequality and in turn gives an upper-bound on the relative error between $z_i$ and $y_i$.
\end{IEEEproof}




\subsection{Design of $K$} \label{sec:controller}



Let us now note that the direct synthesis of a set-point regulator for the hybrid system \eqref{comb} may be difficult. Hence, to solve Problem \ref{problem}-1, we resort to finding sufficient conditions for regulating the hybrid system a desired set-point 
by considering the regulation of an auxiliary uncertain system, 
as stated in the following proposition.


%
   

\begin{prop}
\label{UncertainStab}
The hybrid system \eqref{comb} (with $(A,B,c')$ stabilizable, detectable and $\rho \in (0,1)$) 
can be regulated to an $\epsilon$-neighborhood of a \textit{stabilizable} state $x_d$ via a controller $K$
\noindent if the following auxiliary uncertain system 
\begin{align} \label{uncertain}
\begin{array}{rl}
\dot{x}&=Ax +Bu,\\
z_i&=(1+\Delta_z^i) y_i, \ |\Delta_z^i| \le \delta_z^i, \ \forall i=(1,\dots,r),
\end{array}
\end{align}
can be regulated to an $\epsilon$-neighborhood of the \textit{stabilizable} state $x_d$ via the controller $K$ with $\delta_z^i$ given in Lemma \ref{SectorError}. 
\end{prop}

\begin{IEEEproof}
Lemma \ref{SectorError} shows that the hybrid system \eqref{comb} is an instance of the auxiliary uncertain system \eqref{uncertain}. Thus, the proposition holds directly.
\end{IEEEproof}

Let us now formulate the control problem in view of \eqref{uncertain}. Consider the estimate of the output $z_i$ in \eqref{uncertain} and the unknown output $y_i$ given in \eqref{eq:main}, 
$$ z_i =  (1+\Delta_z^i)y_i = (1+\Delta_z^i) c'(x+x_d + \delta x_i),$$
which in view of a change in coordinates, 
\begin{align} \label{eq:change_coords}
\begin{array}{ll}
\bar{z}_i \rightarrow z_i - c'(x_d + \delta x_i),
\end{array}
\end{align}
becomes
\begin{align} \label{eq:bar_z}
\begin{array}{ll}
\bar{z}_i =  (1 + \Delta_z^i)c'x +  \Delta_z^i c'(x_d + \delta x_i).
\end{array}
\end{align}
Now, the control problem in view of the system dynamics governed by $(A,B)$, the performance variable being the system's state and estimate of the output given by \eqref{eq:bar_z}, which after some algebraic manipulation takes the form,
\begin{align} \label{eq:hinf-prob}
\begin{array}{ll}
\dot{x} &= Ax + Bu, \\
z_p &= C_{z_p}x, \\
\bf{\bar{z}} &=  C_{\bf{\bar{z}}}x + D_{\bf{\bar{z}} \bf{w_1}}{\bf{w_1}} + I_{n \times n} \bf{w_2},
\end{array}
\end{align}
where,
$$ C_{z_p} \triangleq \begin{bmatrix}
I_{n \times n}\\
\vdots \\
I_{n \times n}
\end{bmatrix} \in \mathbb{R}^{nr \times n}, \   C_{\bf{\bar{z}}} \triangleq \begin{bmatrix}
c'\\
\vdots \\
c'
\end{bmatrix} \in \mathbb{R}^{r \times n}, $$
$$D_{\bf{\bar{z}} \bf{w_1}}  \triangleq \begin{bmatrix}
{c'}\\
&{\ddots}\\
&&{c'}\\ 
\end{bmatrix} \in \mathbb{R}^{r \times nr}, \ {\bf{w_1}} \triangleq \boldsymbol{\underline{\Lambda}} z_p \in \mathbb{R}^{nr}, $$
$${\bf{w_2}}\triangleq \boldsymbol{\Lambda} D \in \mathbb{R}^{r} , \  \boldsymbol{\Lambda} \triangleq {\rm diag}({\Delta_z^1, \cdots,\Delta_z^r}) \in \mathbb{R}^{r \times r}, $$
$$ \boldsymbol{\underline{\Lambda}} \triangleq {\rm diag}(I_{n \times n} \otimes \Delta_z^1 , \cdots, I_{n \times n} \otimes \Delta_z^r)   \in \mathbb{R}^{nr \times nr},  $$
$$ {\bf{\bar{z}}} \triangleq \begin{bmatrix}
\bar{z}_1  \\
\vdots \\
\bar{z}_r 
\end{bmatrix}  \in \mathbb{R}^{r}, \ D = \begin{bmatrix}
c'(x_d + \delta x_1)\\
\vdots \\
c'(x_d + \delta x_r)
\end{bmatrix}\in \mathbb{R}^{r}, $$
with ${\bf{w_1}},{\bf{w_2}}$ acting as forcing to the system and $\otimes$ denotes the Kronecker product. It seems natural to put the control problem represented by \eqref{eq:hinf-prob} in the Robust Control framework due to the uncertainty blocks $\boldsymbol{\underline{\Lambda}} ,\boldsymbol{\Lambda}$. More precisely, we wish to utilize the $H_\infty$ controller as it captures the worst case behavior of a system instead of the $H_2$ controller, for instance. In this setting, ${\bf{w_1}}$ represents the component of disturbance whose effect on the performance variable $z_p$ is minimized by virtue of the $H_\infty$ controller. Additionally, ${\bf{w_2}}$ can be thought of as an exogenous (but bounded) disturbance acting on the system. 

\begin{figure}
\centering
    \includegraphics[width=2.5in]{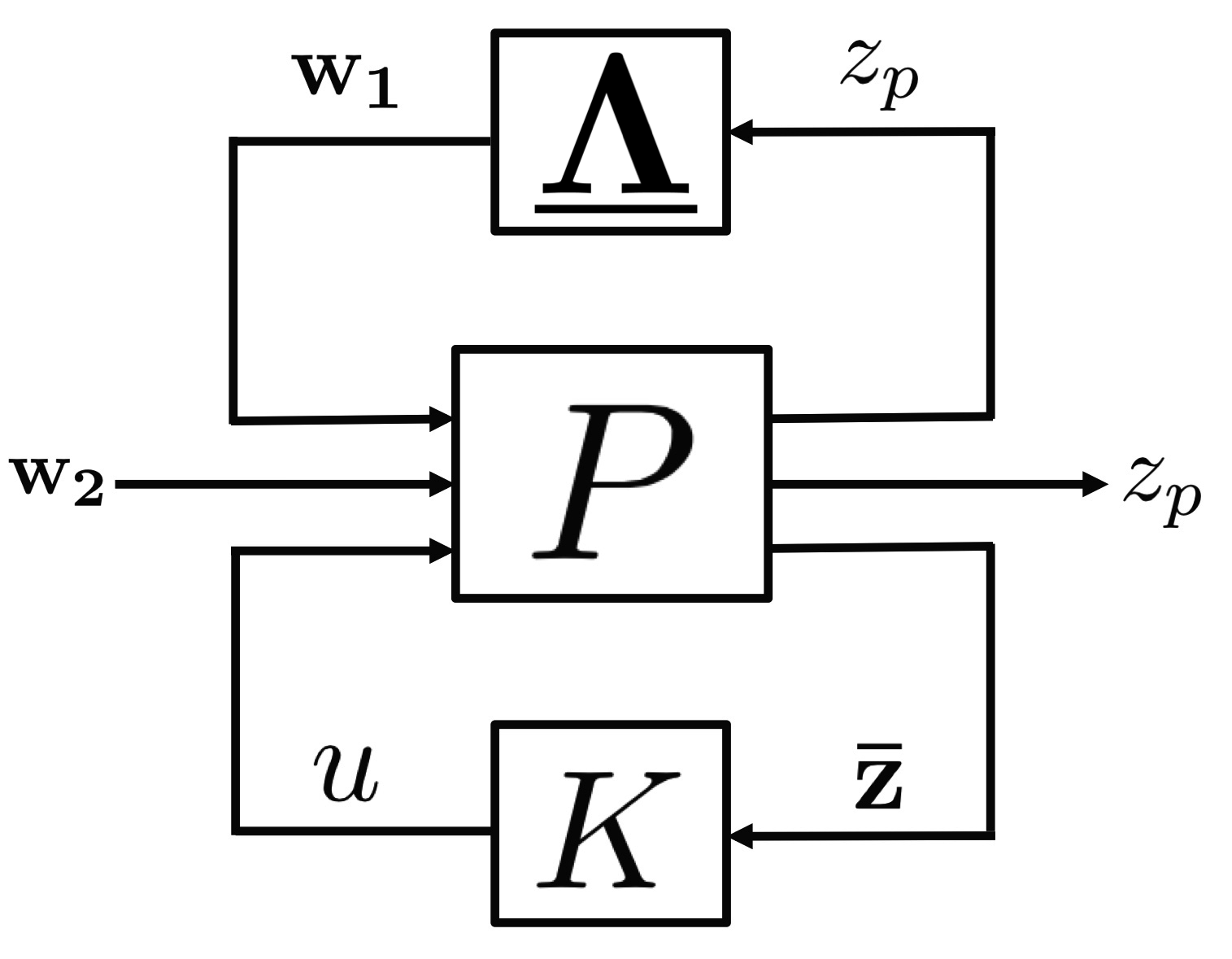}

  \caption{$H_\infty$ control problem with generalized plant $P$ and controller $K$ given in \eqref{eq:P} and \eqref{eq:K}, respectively.   \label{Hinf}}  
\end{figure}

The problem \eqref{eq:hinf-prob} with an additional performance variable $z_p$, for reasons that will become clear later, is illustrated in Figure \ref{Hinf} and the generalized plant $P$ is given by
\begin{align} \label{eq:P}
P=\mleft[ \begin{array}{c|cc} A  & 0_{n\times (nr+r)} & B   \\ \hline C_{z_p} & 0_{nr \times (nr+r)} & \ 0_{nr \times m} \\  C_{z_p} & 0_{nr \times (nr+r)} & \ 0_{nr \times m} \\ C_{\bf{\bar{z}}} & [D_{\bf{\bar{z}} \bf{w_1}} \ I_{r \times r}] & 0_{r \times m} \end{array} \mright]
\end{align} 
and the $H_{\infty}$ controller $K$ given by,
\begin{align}\label{eq:K}
K=\mleft[ \begin{array}{c|c} A_c & B_c  \\ \hline C_c & D_c \end{array} \mright],
\end{align} \\
with input $\bar{z}_i$ given by \eqref{eq:bar_z} and $z_i$ satisfies Lemma \ref{SectorError}, and can be synthesized under some mild assumptions given in \cite{doyle1989state}. Now, the closed loop system arising from the $P$-$K$ combination satisfies
\begin{align} \label{eq:P-K}
\begin{bmatrix}
z_p\\ 
z_p
\end{bmatrix} = \begin{bmatrix}
G_{{\bf{w_1}} z_p} (s) & G_{{\bf{w_2}} z_p} (s) \\ 
G_{{\bf{w_1}} z_p} (s) & G_{{\bf{w_2}} z_p} (s) 
\end{bmatrix} \begin{bmatrix}
{\bf{w_1}}\\ 
{\bf{w_2}}
\end{bmatrix},
\end{align} 
where $G_{{\bf{w_1}} z_p}(s) , G_{{\bf{w_2}} z_p}(s) $ are stable (due to the compensator $K$) transfer functions from disturbance inputs $\bf{w_1},\bf{w_2}$ to the performance variable $z_p$, respectively. These transfer functions are computed, after appropriately combining \eqref{eq:hinf-prob} and the controller \eqref{eq:K}, as
\begin{align*} 
G_{{\bf{w_1}} z_p}(s) &=  C_{z_p} (I^{(n \times n)} - E(s) C_{\bar{z}})^{-1} E(s) D_{\bf{\bar{z}} \bf{w_1}},\\
G_{{\bf{w_2}} z_p}(s) &=  C_{z_p} (I^{(n \times n)} - E(s) C_{\bar{z}})^{-1} E(s),
\end{align*}
where 
$$E(s) \triangleq (sI^{(n \times n)}-A)^{-1}B(C_c (sI^{(n \times n)}-A_c)^{-1}B_c+D_c).$$ 

We are now ready to state the solution to Problem \ref{problem}-1 in the following theorem.

\begin{thm} \label{thm:delta}
The hybrid system \eqref{comb} (with $(A,B,c')$ stabilizable, detectable and $\rho \in (0,1)$) 
can be regulated to an $\epsilon$-neighborhood of a \textit{stabilizable} state $x_d$ (as $t \rightarrow \infty$) 
via an $H_{\infty}$ controller \eqref{eq:K} if $\delta_z^* \triangleq \max_i \delta_z^i$, where 
$\delta_z^i$ is defined in Lemma 4, satisfies
\begin{align} \label{eq:delta}
\delta_z^i \leq \delta_z^* < \frac{\epsilon }{||G_{{\bf{w_2}} z_p}(0)||_2 \frac{||D||_2}{\sqrt{r}} + \epsilon ||G_{{\bf{w_1}} z_p}(0)||_2},
\end{align}

\noindent $(\forall i = 1,\cdots,r)$ where $D$ is defined in \eqref{eq:hinf-prob}, while $G_{{\bf{w_1}} z_p}(s)$ and $G_{{\bf{w_2}} z_p}(s)$ are transfer functions from the respective disturbance components to the performance variable arising from the closed-loop uncertain system \eqref{uncertain} considering measurements $z_i$ from all $r$ pixels.

%
%
\end{thm}

\begin{IEEEproof}
We will show that the auxiliary uncertain system \eqref{uncertain} can be regulated with the $H_{\infty}$ controller \eqref{eq:K} with a proper characterization of the uncertainty in the estimate of the output ${\bf{\bar{z}}}$ in conjunction with the small gain theorem.

In order to meet the objective of Problem \ref{problem}-1, i.e.,
\begin{align}\label{eq:eps-bound} 
\lim_{t \rightarrow \infty} ||x^o(t) - x_d||_2 = \lim_{t \rightarrow \infty} ||x(t)||_2 < \epsilon, 
\end{align}
we will use the final value theorem on the transfer function relating the performance variable $z_p$ to the component of disturbance whose effect has not been taken into account on the system by the $H_\infty$ controller, 
\begin{align}\label{eq:final} 
z_p(\infty) = \lim_{s \rightarrow 0} s \cdot G^{CL}_{{\bf{w_2}} z_p} (s) \frac{\bf{w_2}}{s},
\end{align}
where $G^{CL}_{{\bf{w_2}} z_p} (s)$ takes into account the effect of ${\bf{w_1}} \triangleq \boldsymbol{\underline{\Lambda}} z_p$ in \eqref{eq:P-K} and is found to be 
\begin{align}\label{eq:c-loop} 
G^{CL}_{{\bf{w_2}} z_p} = G_{{\bf{w_2}} z_p} +  G_{{\bf{w_1}} z_p} \boldsymbol{\underline{\Lambda}} (I - G_{{\bf{w_1}} z_p} \boldsymbol{\underline{\Lambda}})^{-1} G_{{\bf{w_2}} z_p}.
\end{align}

However, the final value theorem only applies for stable systems. Although the $H_{\infty}$ controller \eqref{eq:K} will make the transfer matrices $G_{{\bf{w_1}}z_p}$ and $G_{{\bf{w_2}}z_p}$ in \eqref{eq:P-K} stable, we further need to ensure that the closed loop system \eqref{eq:c-loop} is stable in the presence of the uncertainty $\boldsymbol{\underline{\Lambda}}$. To this end, we resort to applying the unstructured small gain theorem to $G_{{\bf{w_1}}z_p}$ that takes the output of the $\boldsymbol{\underline{\Lambda}}$ block as an input
\begin{align} \label{eq:smallgainthm}
 \|G_{{\bf{w_1}}z_p}\|_2 \|\boldsymbol{\underline{\Lambda}} \|_2 <1,
\end{align}
where we will relate $\delta^i_z \leq \delta^*_z \triangleq \|\boldsymbol{\underline{\Lambda}} \|_2$ to the common event threshold $h$ for all pixels in Theorem \ref{thm:hstar}.

Then,  in view of ${\bf{w_2}} \triangleq \boldsymbol{\Lambda} D$, $z_p = C_{z_p}x$, with the application of the induced 2-norm to \eqref{eq:final}, we obtain 
\begin{align}\label{eq:final_x} 
\sqrt{r} ||x(\infty)||_2 = ||z_p(\infty)||_2 \le  ||G^{CL}_{D z_p} (0)||_2 ||D||_2,
\end{align}
where
\begin{align}\label{eq:relation} 
G^{CL}_{D z_p} (s) =  G^{CL}_{{\bf{w_2}} z_p} (s) \boldsymbol{\Lambda} (s).
\end{align} 

By the triangle inequality, from \eqref{eq:c-loop} and \eqref{eq:relation}, we have 
\begin{align*} 
||G^{CL}_{D z_p}||_2  &\leq  ||G_{{\bf{w_2}} z_p} \boldsymbol{\Lambda}||_2\\
 &\quad + ||G_{{\bf{w_1}} z_p} \boldsymbol{\underline{\Lambda}} (I \hspace{-0.05cm}- G_{{\bf{w_1}} z_p} \boldsymbol{\underline{\Lambda}})^{-1} G_{{\bf{w_2}} z_p} \boldsymbol{\Lambda}||_2, 
\end{align*} 
\noindent and by the sub-multiplicative property for induced norms, 
\begin{align}\label{eq:fvt} 
||G^{CL}_{D z_p} (0)||_2 \le \frac{||G_{{\bf{w_2}} z_p}(0)||_2 \delta_z^* }{1-||G_{{\bf{w_1}} z_p} (0)||_2 \delta_z^* }, 
\end{align} 
where we have used the upper-bound for the norm of an inverse operator,
\begin{align}\label{eq:inverse}
  ||(I - G_{{\bf{w_1}} z_p} (0) \boldsymbol{\underline{\Lambda}} (0))^{-1} ||_2 \leq \frac{1}{1 - ||G_{{\bf{w_1}} z_p} (0) \boldsymbol{\underline{\Lambda}} (0)||_2},  
\end{align}
\noindent which holds for 
$$||G_{{\bf{w_1}} z_p} (0) \boldsymbol{\underline{\Lambda}}(0)||_2 \leq ||G_{{\bf{w_1}} z_p} (0)||_2 ||\boldsymbol{\underline{\Lambda}} (0)||_2   < 1,$$ 
by the small gain theorem in \eqref{eq:smallgainthm}. Next, we further upper bound \eqref{eq:inverse} as follows
\begin{align*}
||(I - G_{{\bf{w_1}} z_p} (0)\boldsymbol{\underline{\Lambda}}(0))^{-1} ||_2 &\leq \frac{1}{1 - ||G_{{\bf{w_1}} z_p} (0) \boldsymbol{\underline{\Lambda}}(0)||_2} \\
&\leq \frac{1}{1 - ||G_{{\bf{w_1}} z_p} (0)||_2 \delta_z^* }.
\end{align*} 

Now, upper bounding \eqref{eq:final_x} using \eqref{eq:fvt} and re-arranging gives 
$$ ||x(\infty)||_2 \le \frac{||G_{{\bf{w_2}} z_p}(0)||_2 \delta_z^*}{1-||G_{{\bf{w_1}} z_p} (0)||_2 \delta_z^*} \frac{||D||_2}{\sqrt{r}}, $$ 
and further upper bounding the above in view of \eqref{eq:eps-bound} gives 
$$ ||x(\infty)||_2 \le \frac{||G_{{\bf{w_2}} z_p}(0)||_2 \delta_z^* }{1-||G_{{\bf{w_1}} z_p} (0)||_2 \delta_z^*} \frac{||D||_2}{\sqrt{r}} < \epsilon, $$ 
and upon re-arranging, we obtain the expression in \eqref{eq:delta} for the maximum uncertainty that can be tolerated. Note that \eqref{eq:delta} (with $D=0$) implies that the small gain theorem in \eqref{eq:smallgainthm} holds, which in turn implies that \eqref{eq:delta} is sufficient for closed-loop stability (such that the final value theorem applies) and for the regulation of 
the closed loop system $(P,K,\Delta^i_z)$ $\forall i=(1,\dots,r)$ to an $\epsilon$-neighborhood of a \textit{stabilizable} state $x_d$ (as $t \rightarrow \infty$). 
Finally, by Proposition 1, the $H_{\infty}$ controller asymptotically regulates the the hybrid system \eqref{comb} to an $\epsilon$-neighborhood of a \textit{stabilizable} state $x_d$.
%
%
%
\end{IEEEproof}


We now solve Problem \ref{problem}-2 in the following theorem.

\begin{thm} \label{thm:hstar}
The hybrid system \eqref{comb} (with $(A,B,c')$ stabilizable, detectable and $\rho \in (0,1)$) 
can be regulated to an $\epsilon$-neighborhood of a \textit{stabilizable} state $x_d$ (as $t \rightarrow \infty$) 
via an $H_{\infty}$ controller \eqref{eq:K} provided that the event threshold $h$ in \eqref{eq:trigger} for the DVS is upper-bounded by
\begin{align} \label{eq:hstar}
h \leq h^\star = \log_b \sqrt{\min_i \left(  \frac{m_i}{M_i} \right)\frac{1 +  \delta_z^*}{1 -  \delta_z^*}},
\end{align}
with $\delta_z^i,\delta_z^*$ as defined in Lemma \ref{SectorError} and Theorem \ref{thm:delta}, respectively. 
\end{thm}

\begin{IEEEproof} 
We assume that there is one common event threshold $h$ (and correspondingly, $\rho$) for all pixels and we would like to use each pixel for the regulation task. Hence, in view of the expression for $\rho$ in Lemma \ref{SectorError}, we choose $\rho^*$ that corresponds to the smallest $h^*$ ($\rho^* \propto \frac{1}{h^*}$ according to \eqref{eq:rho}), i.e.,
$$\rho^* = \sqrt{\max_i \left(\frac{M_i}{m_i}\right)\frac{1 -  \delta_z^*}{1 +  \delta_z^*}}.$$
Then, from our definition of $\rho$ in \eqref{eq:rho}, we obtain the expression given by \eqref{eq:hstar}. Finally, $h^*$ is an upper bound on the tolerable event threshold $h$ since it can be verified that any $h \leq h^*$ implies that $\delta_z^i \leq \delta_z^*$ and thus \eqref{eq:delta} holds.  
\end{IEEEproof}

\begin{rem}\label{rem:cases}
In Section \ref{sec:num}, we shall compare the performance of the closed loop hybrid system \eqref{comb} by guaranteeing that Theorem \ref{thm:hstar} holds and by synthesizing a robust controller $K$ for the uncertain open-loop plant model \eqref{eq:P} considering the uncertainties, $\boldsymbol{\underline{\Lambda}},\boldsymbol{\Lambda}$, to be of
\begin{enumerate}
\item an unstructured form as in \cite{doyle1989state}, and
  \item a structured form where we will employ the $D$-$K$ algorithm for $\mu$-synthesis as in \cite{packard1993linear}.
\end{enumerate}
\end{rem}

In Lemma \ref{SectorError}, we reasoned that the minimization of the relative error between $z_i$ and $y_i$ 
would be obtained by having symmetric error bounds on $z_i$ via a choice of $\lambda_i = \frac{(M_i +m_i) \rho}{M_i +m_i \rho^2}$. However, this reasoning needs verification. 
For the $H_{\infty}$ controller \eqref{eq:K} that satisfies Theorem \ref{thm:delta}, we verify in the following Theorem that the threshold $h^*$ in Theorem \ref{thm:hstar} indeed solves Problem \ref{problem}-2. In turn, we also verify the claim in Remark \ref{estimator_choice}.


\begin{thm} \label{thm:lambda}
The choices of $\hat{q}_i (0) = \frac{2m_iM_i}{M_i+m_i}$ and  $\lambda_i = \frac{(M_i +m_i) \rho}{M_i +m_i \rho^2}$ in Lemmas \ref{qhat-estimator-lem2} and \ref{SectorError}, respectively, yield the least restrictive (largest) upper bound on the event threshold $h^*$ in Theorem \ref{thm:hstar}. 
\end{thm}
\begin{IEEEproof}
The problem of finding the least restrictive upper bound on the event-threshold, $h^*$, in \eqref{eq:hstar} for the DVS is equivalent to finding the minimum $\rho^*$, and can be cast, for each pixel $i$, as the following optimization problem (dependence on $i$ is omitted for simplicity):
\begin{align}
\begin{array}{rl}
\underset{\rho, \lambda,\hat{q}(0)}{\text{minimize}} \quad & \rho, \\
\text{subject to} \quad & 0<\rho<1, \\
& \delta_z = \textrm{max} \{1-\lambda \hat{q}(0) \frac{\rho}{M},\frac{\lambda \hat{q}(0)}{m\rho} -1\}, \\
& 0 < \delta_z \le \bar{\delta},
\end{array} \label{eq:opt1}
\end{align}
with a $\bar{\delta}$ satisfying $\bar{\delta}< \delta_z^*$ as defined in Theorem \ref{thm:delta}. To obtain the equality constraint in \eqref{eq:opt1}, we first employ the lower and upper bounds of $\Delta_q$ given in \eqref{q_vals_bound} into \eqref{bounds_del}, thus, giving us 
$$\left(1+\frac{\hat{q}(0)-M}{M}\right) \rho y  \leq \hat{q}  \leq \left(1+\frac{\hat{q}(0)-m}{m}\right) \frac{y}{\rho}. $$
In view of \eqref{y-observer}, the above becomes 
$$\lambda \hat{q}(0)\frac{\rho}{M} -1 \leq \frac{z-y}{y} \leq \frac{\lambda \hat{q}(0)}{m\rho} - 1.$$
Then, defining $\delta_z$ as the upper bound on $|\frac{z-y}{y} |$, the equality constraint in \eqref{eq:opt1} follows. Now, defining a new decision variable $\bar{\lambda} \triangleq \lambda \hat{q}(0)$, the above optimization problem can be solved analytically by noting that $\rho^*$ is given by $\min\{\rho_a,\rho_b\}$ in view of the following two sub-problems (cf. Figure \ref{best_HM}): \vskip-0.75cm
\begin{align*}
\begin{split}
& \underset{\rho_a, \bar{\lambda}_a}{\text{minimize}} \ \rho_a, \\
& {\text{subject to:}} \\ 
& \hspace{0.5cm} 0<\rho_a<1, \\
& \hspace{0.5cm} \frac{2mM \rho_a}{M + m \rho_a^2} \leq \bar{\lambda}_a, \\
& \hspace{0.5cm} m \rho_a \hspace{-0.05cm}< \hspace{-0.05cm}{\bar{\lambda}_a}  \hspace{-0.05cm}\le \hspace{-0.05cm} m\rho_a(\bar{\delta} \hspace{-0.1cm}+\hspace{-0.05cm}1),
\end{split} 
\begin{split}
\\[-1.ex]
& \underset{\rho_b, \bar{\lambda}_b}{\text{minimize}} \ \rho_b, \\
& {\text{subject to:}} \\  
& \hspace{0.5cm} 0<\rho_b<1, \\
& \hspace{0.5cm} \frac{2mM \rho_b}{M + m \rho_b^2} \geq \bar{\lambda}_b, \\
& \hspace{0.5cm} \frac{M(1-\bar{\delta})}{\rho_b}  \le \hspace{-0.05cm} \bar{\lambda}_b  \hspace{-0.05cm} < \hspace{-0.05cm} \frac{M}{\rho_b}.
\end{split}
\end{align*}
\begin{figure}
\centering
  \includegraphics[width=3.4in,trim= 10mm 2mm 2mm 10mm]{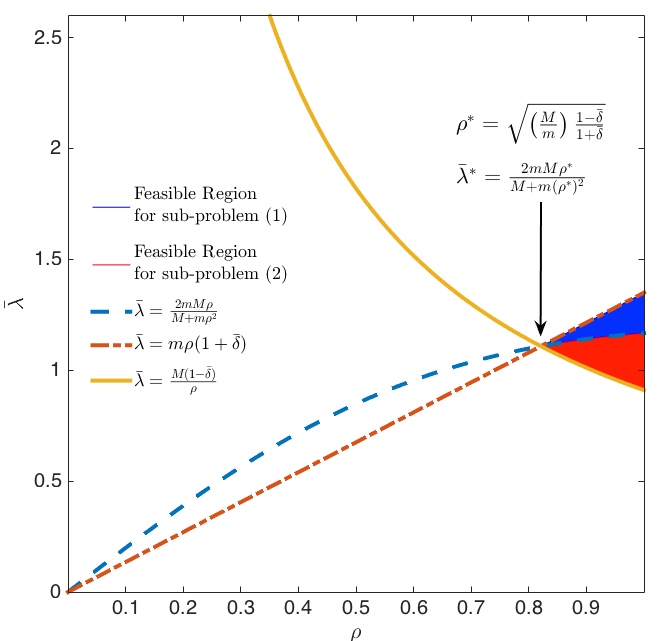}
  \caption{Illustration of the optimization problem for finding $(\rho^*,\bar{\lambda}^*)$ via two sub-problems (with $\bar{\delta}=0.35,m=1,M=1.4$).}
  \label{best_HM}
\end{figure}

It can be verified that the solutions to both sub-problems coincide in a unique $\bar{\lambda}^*=\bar{\lambda}(\rho^*)$, as illustrated in Figure \ref{best_HM}. Thus, this concludes the proof since $\rho^*$ results in the least restrictive upper bound $h^*=- \log_b \rho^*$ in view of \eqref{eq:rho} and $\lambda \hat{q}(0)= \bar{\lambda}^*$  is a particular solution to the optimization problem.
\end{IEEEproof}

\begin{rem}\label{rem:one_param}
The ability to capture two decision variables within one, $\bar{\lambda}_i \triangleq \lambda_i \hat{q}_i(0)$, in Theorem \ref{thm:lambda}, follows from the fact that in view of \eqref{q-observer}, the evolution of $\hat{q}_i$ follows 
$$\hat{q}_i = \hat{q}_i(0) \rho^{-\sum_i p_i}. $$
Thus, $z_i$ in \eqref{y-observer} becomes
$$z_i = \lambda_i \hat{q}_i =  \lambda_i \hat{q}_i(0) \rho^{-\sum_i p_i} = \bar{\lambda}_i \rho^{-\sum_i p_i},$$
which reveals that there exists an infinite number of choices for $\lambda_i$ and $\hat{q}_i(0)$ such that $\lambda_i \hat{q}_i(0)$ equals the unique  $\bar{\lambda}^*$. 
Mathematically, this realization suggests that one need not design $\hat{q}_i(0)$ and then $\lambda_i$ for the $i$'th pixel to meet the regulation task, i.e., using the procedure outlined in Figures \ref{feedback},\ref{wfunc}. Instead one can choose any design route that satisfies the constraint,
$$ \bar{\lambda}_i \triangleq \lambda \hat{q}(0) = \frac{2m_iM_i \rho}{M_i + m_i \rho^2},$$
to fulfill the regulation task.
\end{rem}

\begin{rem}\label{rem:HM_coll}
It is relatively clear that if $q(0)$ is known (i.e., $q(0)=m=M$), then, $\lambda_i = \lambda_{HM} = \frac{2\rho}{1+\rho^2}$, which has been shown in \cite{singh.yong.ea.2016}.
\end{rem}

\begin{figure}[t]
\centering
\includegraphics[width=1\linewidth]{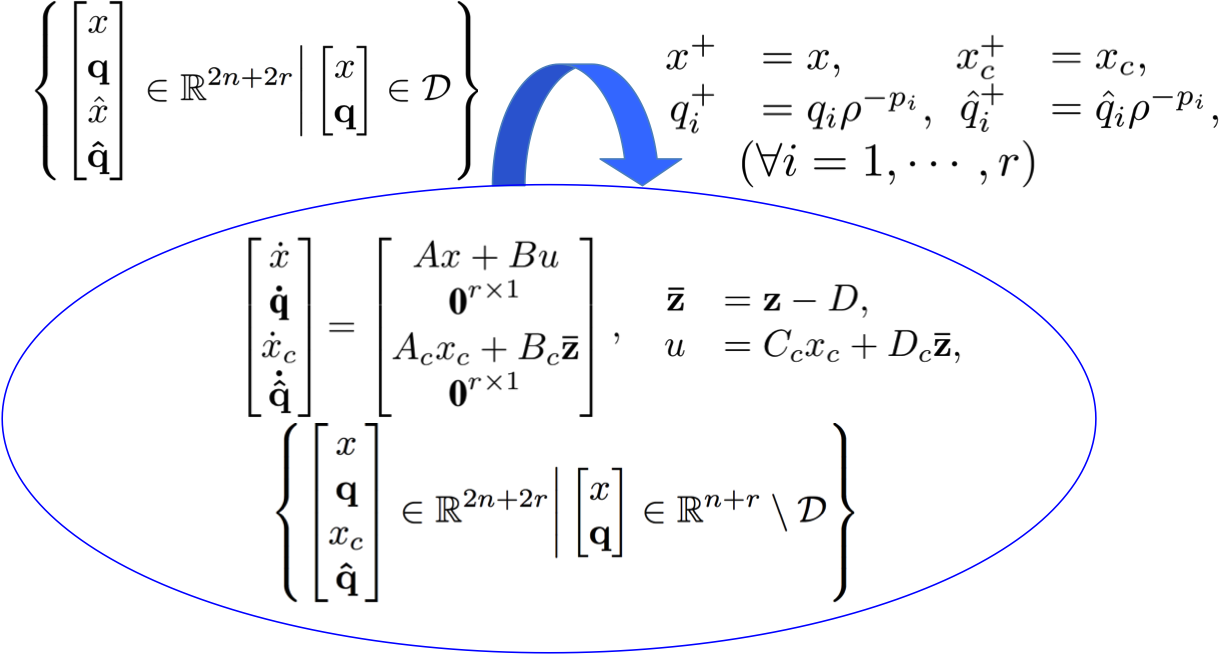} \caption{Closed loop hybrid automaton of combined LTI system, DVS model and $H_{\infty}$ controller, where $\mathcal{D} = \cup_{i=1}^r\mathcal{D}_i$ and $\mathcal{D}_i$ is defined in \eqref{eq:guardSet}, the solution of the continuous dynamics (i.e., in between events) lies in $ \mathbb R^{2n+2r} \textbackslash
\mathcal D $,
$\textbf{z} = [z_1(t),\dots,z_r(t)]^T$ with $z_i$ and $D$ defined in \eqref{y-observer}, \eqref{eq:hinf-prob} respectively. \label{fig:hybridAutomaton2}}
\end{figure}


\begin{rem}\label{rem:practical}
A potential problem that can arise in the practical implementation of our approach when regulating to the origin, i.e., $D=0$ is the possibility of having an infinite number of events when the output is near the origin. This can occur due to the logarithmic \textit{trigger condition} in \eqref{eq:trigger}, which induces logarithmic spacings separated by $\rho$ in the output space that carry over to the state-space due to the linearity of the output $y_i$. In particular, when the output $y_i$ crosses the origin, an infinite number of events would be fired by the DVS over a finite time interval (similar to a Zeno phenomenon observed in hybrid systems, e.g., in \cite{goebel.sanfelice.ea.09}), making it impractical for any  controller to keep up with in real time. 

To overcome this potential problem, we propose the inclusion of an auxiliary band with width $10^{-4}$ near the origin as in \cite{picasso2008hypercubes}. We remark that this problem would not happen for vision sensors as luminosity is nonnegative.
\end{rem}



To sum up, our resulting closed loop hybrid automaton is illustrated in Figure \ref{fig:hybridAutomaton2}.

\section{Numerical Experiment} \label{sec:num}

In this section, we demonstrate our proposed approach outlined in Section \ref{sec:controller} with the following unstable, but stabilizable and detectable, system: 
$$A = \begin{bmatrix}
2 & 10 \\ 
0 & 5
\end{bmatrix}, \ \ B = \begin{bmatrix}
1 \\ 
1
\end{bmatrix}, \ \ c = \frac{1}{\sqrt{5}}\begin{bmatrix}
2 \\ 
1
\end{bmatrix}. $$
In view of Remark \ref{rem:cases}, we consider the task of regulating the system \eqref{comb} from the initial state $x^o(0) \triangleq x_0 =\begin{bmatrix} 0.0179 \\ 0.3428 \end{bmatrix}$ to within 
an $\epsilon$-neighborhood (i.e., tolerance level $\epsilon = 0.05$) of the \textit{stabilizable} state, 
$x_d =  \begin{bmatrix}
-0.2321 \\ 
0.0928
\end{bmatrix}$, 
which has been found following the description below \eqref{eq:coord}, using measurements from nine pixels with relative locations given by
$$\delta x_1 =\begin{bmatrix}
0 \\ 
0
\end{bmatrix} , \ \delta x_2 = \begin{bmatrix}
0.01 \\ 
0
\end{bmatrix}, \ \delta x_3 =  \begin{bmatrix}
-0.01 \\ 
0
\end{bmatrix},$$
$$\delta x_4 =\begin{bmatrix}
-0.01 \\ 
0.01
\end{bmatrix} , \ \delta x_5 = \begin{bmatrix}
0 \\ 
0.01
\end{bmatrix}, \ \delta x_6 =  \begin{bmatrix}
0.01 \\ 
0.01
\end{bmatrix},$$  
$$\delta x_7 =\begin{bmatrix}
0.01 \\ 
-0.01
\end{bmatrix} , \ \delta x_8 = \begin{bmatrix}
0 \\ 
-0.01
\end{bmatrix}, \ \delta x_9 =  \begin{bmatrix}
-0.01 \\ 
-0.01
\end{bmatrix},$$ 
as illustrated in Figure \ref{fig:grid}. We further assume that $m_i = c'(x_0+\delta x_i)-\delta y$ and $M_i=c'(x_0+\delta x_i)+\delta y$ for all $i=\{1,\dots,9\}$ with $\delta y = 0.002$.  Then, the initial values of the estimates of the base of the trigger reference $\hat{q}_i(0)$ can be computed in view of Lemma \ref{qhat-estimator-lem2}. 

The closed loop system \eqref{eq:main1} in the original coordinate frame $x^o$ can now be simulated via the control in the original coordinate $u^o$ given in \eqref{eq:control_coord} under the action of the synthesized controller $u = K {\bf{\bar{z}}}$ discussed in Figure \ref{fig:hybridAutomaton2}.


\begin{figure*}[tph]
\begin{centering}
  \subfigure[Plot of error between the system state and \textit{stabilizable} state, $||x^o(t)-x_d||$.]{
  \includegraphics[width=4.5in]{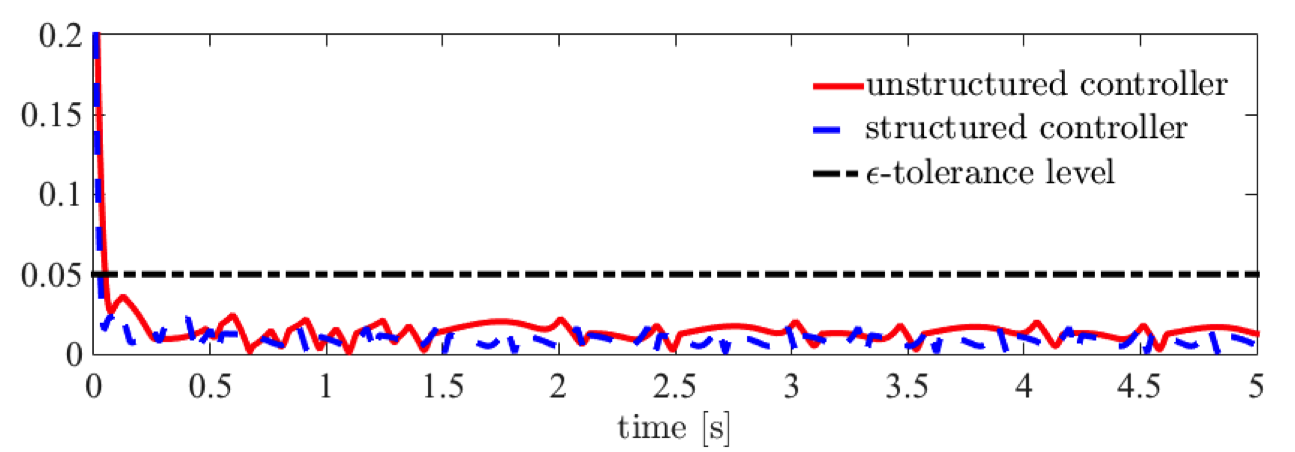}  \label{one_and_three}}

  \subfigure[Phase portrait starting from $x^o(0)=x_0$.]{
  \includegraphics[width=3.25in]{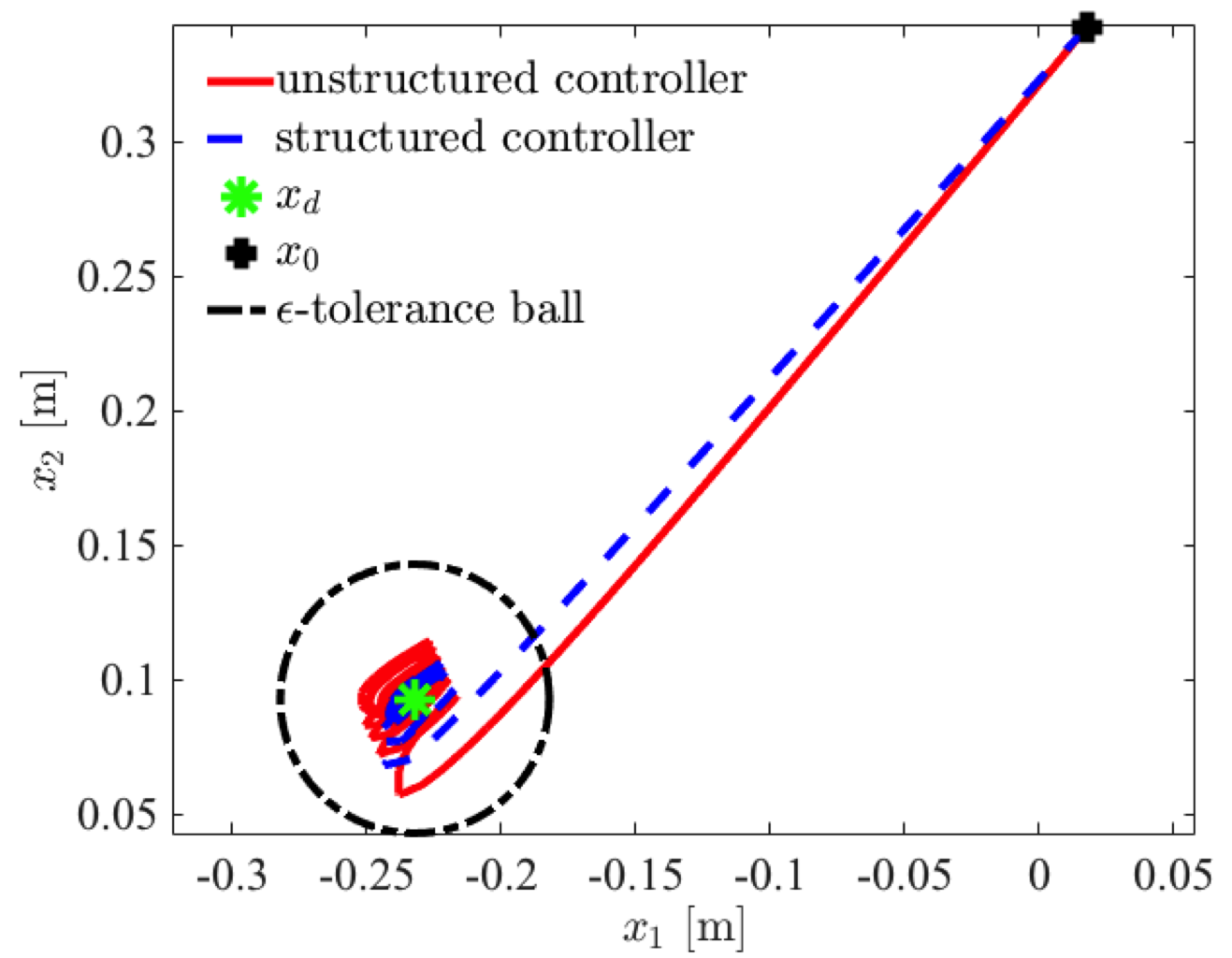}      \label{phase}}
    
  \caption{Closed loop response of system with an $\epsilon = 0.05$ tolerance level using both unstructured and structured controllers with three measurements $(r = 3)$.}
  \end{centering}
\end{figure*}

\subsection{Unstructured vs. structured controller with three measurements ($r=3$)} \label{sec:un_struct}

We synthesize the controller $K$ for the cases of $\boldsymbol{\underline{\Lambda}},\boldsymbol{\Lambda}$ being unstructured and then structured to discuss the performance of the closed-loop system under the action of the respective controllers that process ``retinal events" arriving from three pixels with relative distances $\delta x_1, \delta x_2, \delta x_3$. For the case of $\boldsymbol{\underline{\Lambda}},\boldsymbol{\Lambda}$ being unstructured, we use the \texttt{hinfric} command in MATLAB  to obtain the corresponding $H_{\infty}$ controller $K$ for the auxiliary uncertain system \eqref{eq:hinf-prob} with the second row in \eqref{eq:P} removed. 
The controller for the case of structured $\boldsymbol{\underline{\Lambda}},\boldsymbol{\Lambda}$ for the generalized plant $P$ given in \eqref{eq:P} is synthesized via a two step procedure:

\begin{enumerate}
\item the \texttt{hinfric} command in MATLAB is used to obtain a regularized generalized plant as in the unstructured case, and 
\item the \texttt{dksyn} command in MATLAB is used to generate the controller for the regularized plant and takes into account the structured nature of the uncertainty. 
\end{enumerate}

\begin{rem}\label{rem:}
Note that the feedthrough term relating $u \rightarrow z_p$ in \eqref{eq:hinf-prob} is zero; hence, a regularized plant is required to synthesize an $H_\infty$ controller. The  \texttt{hinfric} command in MATLAB performs this regularization. Thus, for the structured controller, we take advantage of this command in step (1) above to obtain a regularized plant for use in step (2) above. 
\end{rem}


Figure \ref{one_and_three} shows a plot of the evolution of the error between the system's state $x^o(t)$ and the stabilizable state $x_d$ under the action of both the unstructured and structured controllers. The better performance of the structured controller (over the unstructured controller) is clear from the faster rate of decay of the error seen initially and furthermore, the long term average error resulting from the structured controller appears to be slightly better than the long term average error from the unstructured controller. On the other hand, Figure \ref{phase} illustrates the phase portrait of the closed-loop responses $(x^o_1 , x^o_2)$ using both the unstructured and structured controllers. We observe that the deviation of the state trajectory from the desired state is smaller for the structured controller than the unstructured controller. 

\begin{figure*}[tph]
\centering
\includegraphics[width=6.75in]{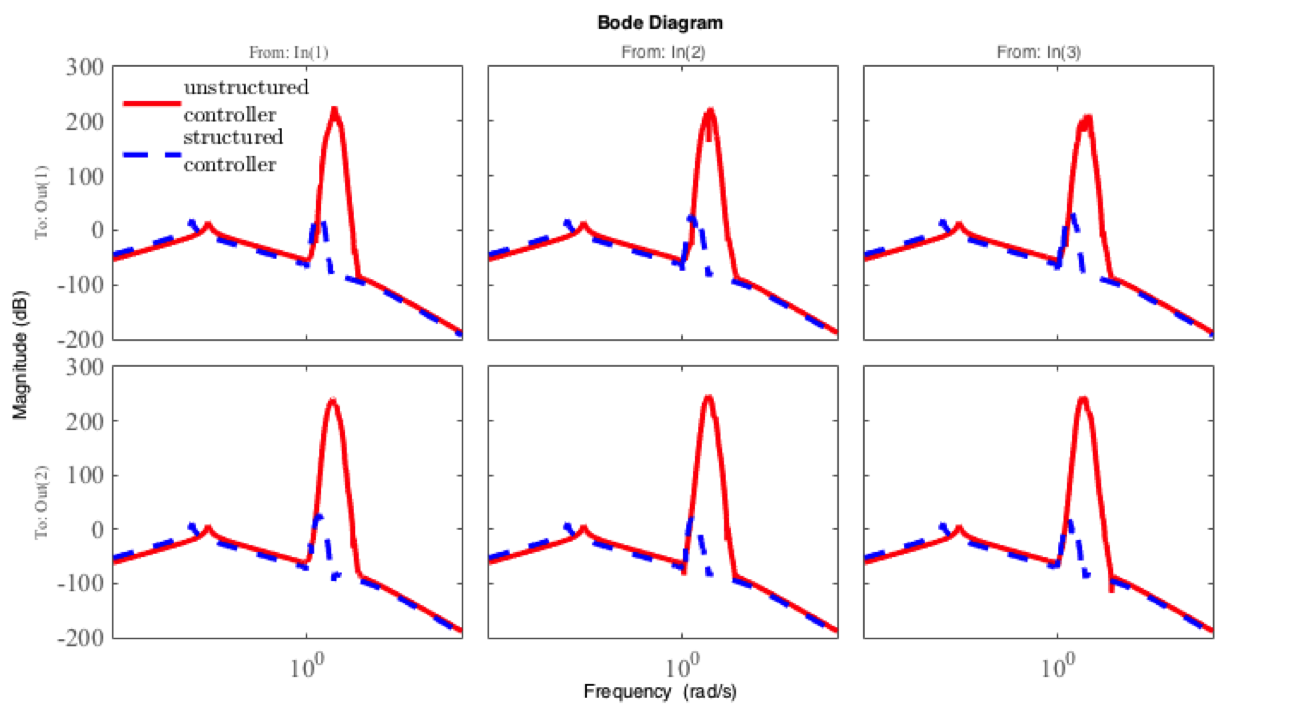} 
\caption{Bode magnitude plot of worst-case performance of closed loop uncertain system $G^{CL}_{D x} (s)$, found from \eqref{eq:relation} in view of $z_p = C_{z_p}x$, using both unstructured and structured controllers with three measurements (i.e., represented as inputs in the plot).}
 \label{bode}
 \end{figure*}

Additionally, from the Bode plot in Figure \ref{bode}, we observe that  
%
the structured controller better attenuates the ``peak" responses at the worst case frequencies than the unstructured controller. In order to do this, the structured controller requires a ``slightly" higher DC gain than the unstructured controller, in effect, constituting to the classical ``waterbed effect" described by Bode's sensitivity integral. The ``slightly" higher DC gain resulting from the structured controller, however, is the reason for the marginally different event thresholds corresponding to the unstructured controller, $h^*_u = 0.1562$, and the structured controller $h^*_s = 0.1510$.
Moreover, as expected, the structured controller better meets the usual control performance specifications, namely, good tracking occurring at low frequencies and good disturbance rejection occurring at high frequencies. 

 \begin{figure*}[tph]
\begin{centering}
  \subfigure[Plot of error between the system state and \textit{stabilizable} state, $||x^o(t)-x_d||$.]{
  \includegraphics[width=4.85in]{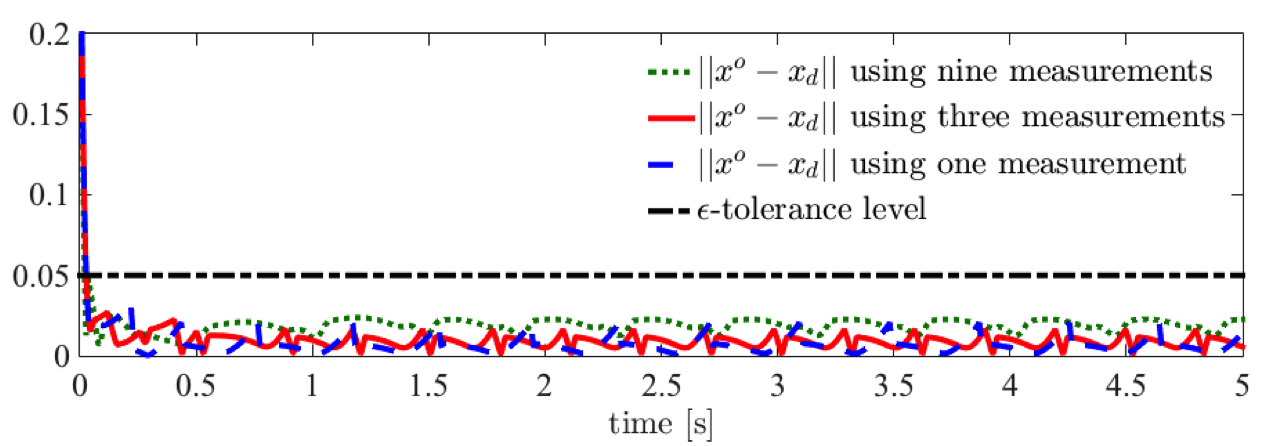}  \label{one_and_three1}}

  \subfigure[Phase portrait starting from $x^o(0)=x_0$.]{
  \includegraphics[width=3.75in]{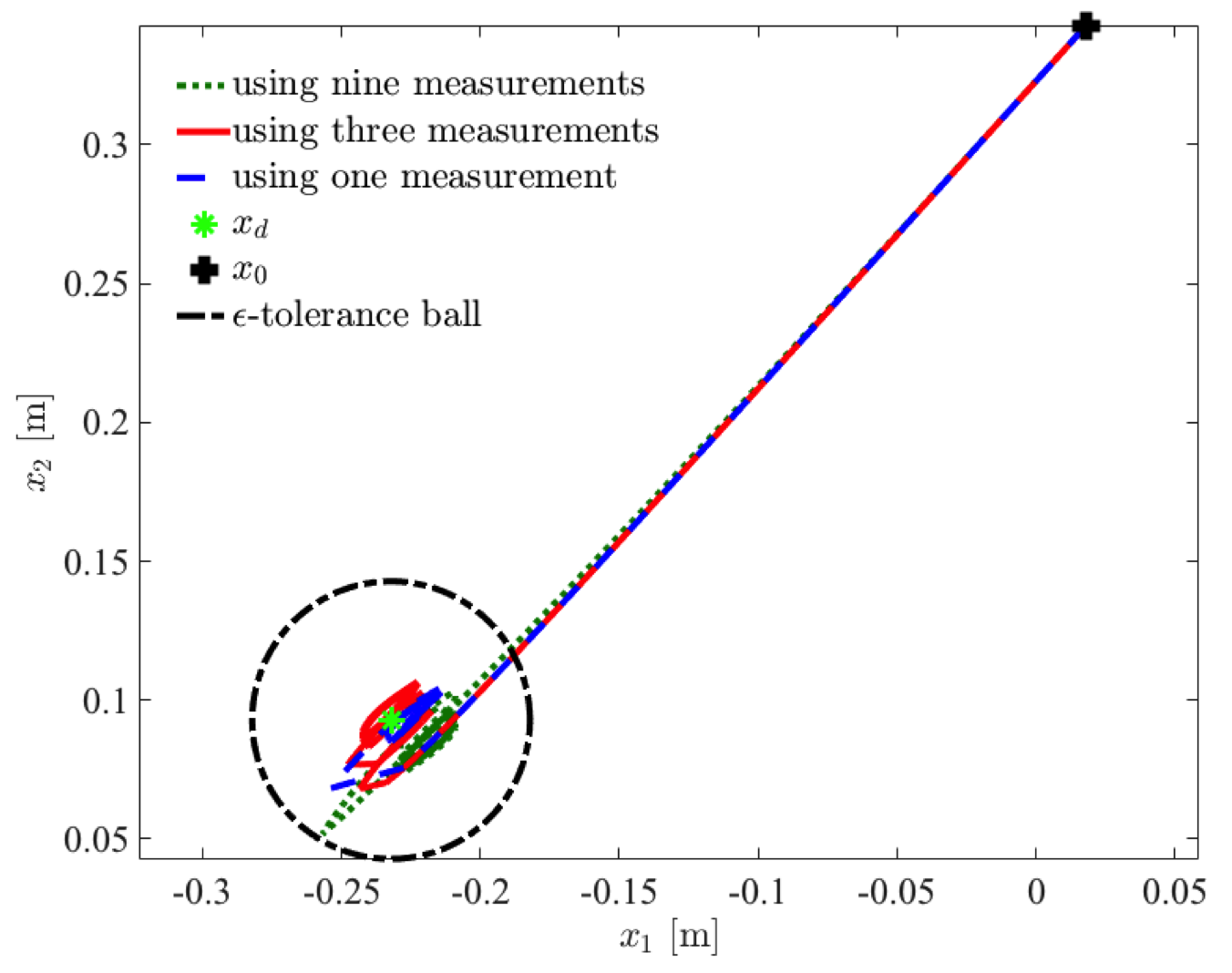}      \label{phase1}}
    
  \caption{Closed loop response of system with an $\epsilon = 0.05$ tolerance level using structured controller for one, three and nine measurements $(r = 1,3,9)$.}
  \end{centering}
\end{figure*}

\subsection{Structured controller processing single and multiple measurements} \label{sec:one_many}

Now, having established, through the previous discussion, that the structured controller yields better closed-loop performance, we consider next the performance of using the structured controller for different numbers of measurements arriving from the DVS. Specifically, we now discuss the performance of the closed-loop system under the action of the structured controller $K$ for the cases of $r = \{1, 3, 9 \}$ measurements arriving from the DVS. We have used $\delta x_1$ for the $r = 1$ case, $(\delta x_1,\delta x_2,\delta x_3)$ for the $r = 3$ case and $(\delta x_1,\dots,\delta x_9)$ for the $r = 9$ case. \\

Figures \ref{one_and_three1} and \ref{phase1} show plots of the evolution of the error between the system's state $x^o(t)$ and the stabilizable state $x_d$ and of the phase portrait of of the closed-loop responses $x^o_1 , x^o_2$, respectively, under the action of the structured controller for the different
numbers of measurements considered. We observe that the long term average error seems to increase with the number of measurements $r$. 
Additionally, the event thresholds $h^*$ for $r = \{1, 3, 9\}$ cases were found to be $\{0.1511, 0.1510, 0.1508\}$, respectively. Both these \emph{observations} seem counter-intuitive as one would hope that more measurements would enable the controller to yield a better closed-loop performance (i.e., smaller long term average error and larger event threshold). This is actually not surprising because of the inverse relation of $||D||_2$ with $\delta_z^*$ in \eqref{eq:delta}, which is then proportional to $h^*$ in \eqref{eq:hstar}. More precisely, it is clear that $\delta x_i$, $\forall i = \{1,\dots,9\}$, having non-zero entries contributes to a larger exogenous disturbance $||D||_2$ (thus, smaller event-threshold) with respect to the increasing number of pixels. Moreover, we have insisted in Theorem \ref{thm:hstar} that we will use each pixel for the regulation task. This means that we are choosing the common event threshold $h^*$ based on the pixel that required the most events, i.e., the smallest threshold. In other words, any increase in the number of pixels will not decrease the threshold from the one corresponding to the `worst' existing pixel.  
Nonetheless, having more measurements may be useful when the measurements are corrupted by stochastic noise, which is part of our future work. 

%

\section{Conclusions and Future Work} 
The Dynamic Vision Sensor (DVS) is a neuromorphic sensor, which is a recent addition to the class of vision sensors. The nice properties of the DVS promise to facilitate agile robotic maneuvers. 
However, existing vision algorithms cannot be directly adapted to process these events; thus, new algorithms need to be developed. 

In this work, we proposed an $H_\infty$ controller that regulates  stabilizable and detectable LTI systems using DVS measurements to a \textit{stabilizable} state $x_d$. In particular, we provide the least restrictive upper bound on the event threshold, $h^*$, for the DVS such that the pair $(A,B)$  can be regulated to within a pre-set tolerance of the \textit{stabilizable} state. This work can be viewed as an initial attempt to locally regulate a nonlinear system about some operating point using DVS measurements. 

There are many interesting directions of future research. An important one being to implement the theory developed here on a practical testbed. Additionally, it is crucial to develop a control scheme that can stabilize a given LTI system in the presence of stochastic noise, 
which is frequently encountered in practice. To conclude, a linear varying luminance profile may not be regularly encountered in practice and so a control scheme needs to be developed that can handle an accurate description of the environment's luminance, e.g., through an integrative sensor model.




\balance

\section*{Acknowledgments}
This work was supported by the Singapore National Research Foundation through the SMART Future Urban Mobility project.

\bibliographystyle{unsrt}


\bibliography{biblio}
\appendices \section{Physical example}  \label{sec:physicalEx}

\begin{figure}[!hbp]
\centering
  \includegraphics[width=1.75in]{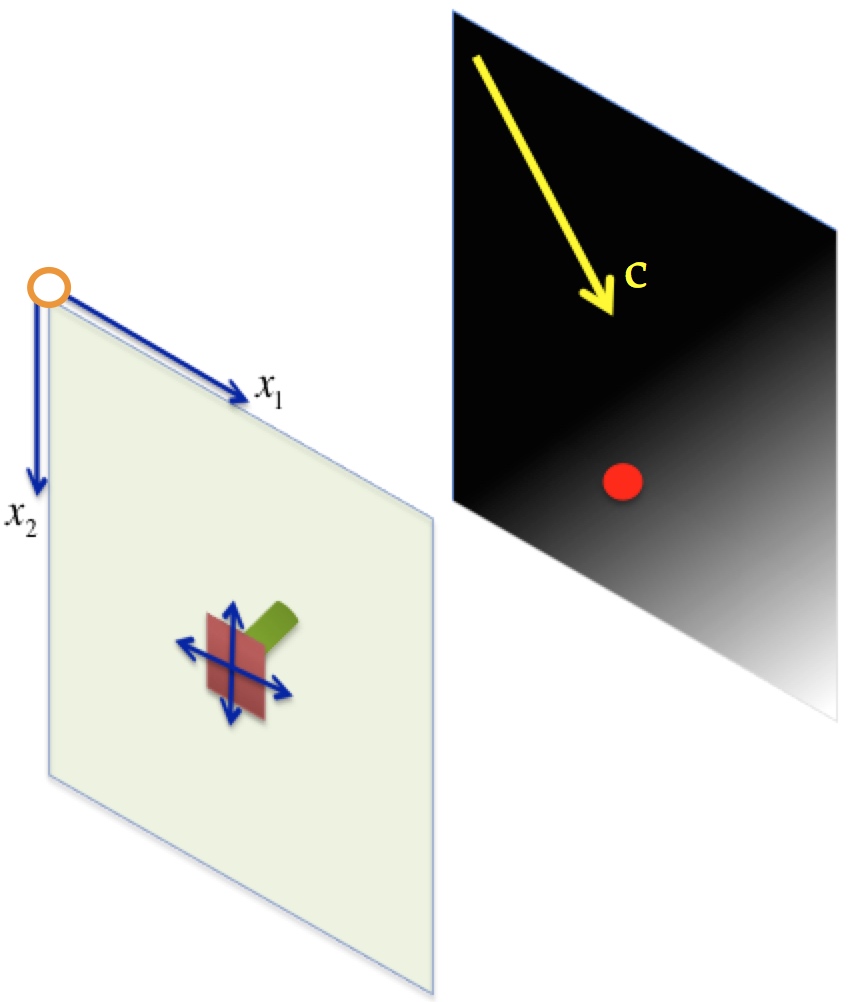}
  \caption{Physical example: DVS (green cylinder) mounted on a platform with linear $(x_1,x_2)$ dynamics and facing a linearly varying brightness profile. The DVS provides control commands $u$ to the platform to move the camera towards the red/solid dot (i.e., the \textit{stabilizable} state).}
  \label{phys}
\end{figure}

Figure \ref{phys} presents a physical example  that may be encountered in practice. In this example, the DVS is mounted on a platform with linear $(x_1,x_2)$ dynamics and looks at a linearly varying brightness profile whose gradient is given by the linearized sensor function, $c$. The $H_{\infty}$ controller developed in Section \ref{sec:controller} produces suitable control commands $u$ to the platform for moving the DVS to the \textit{stabilizable} state. 

\end{document}